\documentclass[12pt]{article}

\usepackage[T1]{fontenc}      
\usepackage[utf8]{inputenc}  
  \usepackage{tgtermes}         

\usepackage{geometry}                
\usepackage{graphicx}
\usepackage{longtable}
\usepackage{natbib}
\usepackage{color}
\usepackage{amssymb}
\usepackage{amsmath}
\usepackage{epstopdf}
\usepackage{boxedminipage}
\usepackage{authblk}
\usepackage{booktabs}
\usepackage{topcapt}
\usepackage{lineno}

\newcounter{reaction}

\title{Creation and Evolution of Impact-generated Reduced Atmospheres of Early Earth}
\author[1]{Kevin J.\ Zahnle}
\author[2]{Roxana Lupu}
\author[3]{David C.\ Catling}
\author[3]{Nick Wogan}

\affil[1]{Space Science Division,
 NASA Ames Research Center,
 Mail Stop 245-3, Moffett Field CA 94035 USA,
 1-650-604-0840,
 Kevin.J.Zahnle@NASA.gov,
{ORCID~ID:~0000-0002-2462-4358}}
\affil[2]{SETI Institute, NASA Ames Research Center, Moffett Field, CA 94035 USA,
Roxana.E.Lupu@nasa.gov
}
\affil[3]{Department of Geology, University of Washington, Seattle, WA 98195 USA,
dcatling@u.washington.edu }
\date{\it Submitted 31 December, 2019; Accepted 10 February 2020} 

\begin{document}
\pagenumbering{roman}
\maketitle

\newpage


\begin{abstract}

  The origin of life on Earth seems to demand a highly reduced early atmosphere, rich in CH$_4$, H$_2$, and NH$_3$, 
 but geological evidence suggests that Earth's mantle has always been relatively oxidized
 and its emissions dominated by CO$_2$, H$_2$O, and N$_2$. 
 The paradox can be resolved by exploiting the reducing power inherent in the ``late veneer,'' i.e., material
 accreted by Earth after the Moon-forming impact.
   Isotopic evidence indicates that the late veneer consisted of extremely dry, highly reduced inner solar system materials, suggesting
 that Earth's oceans were already present when the late veneer came.
 The major primary product of reaction between the late veneer's iron and Earth's water was H$_2$.
  Ocean vaporizing impacts generate high pressures and long cooling times that favor CH$_4$ and NH$_3$. 
  Impacts too small to vaporize the oceans are much less productive of CH$_4$ and NH$_3$, 
 unless (i) catalysts were available to speed their formation, 
 or (ii) additional reducing power was extracted from pre-existing crustal or mantle materials.
 The transient H$_{\rm 2}$-CH$_{\rm 4}$ atmospheres evolve photochemically
 to generate nitrogenated hydrocarbons at rates determined by solar radiation and hydrogen escape,
 on timescales ranging up to tens of millions of years and with cumulative organic production ranging up to half a kilometer. 
 Roughly one ocean of hydrogen escapes. 
 The atmosphere after the methane's gone is typically H$_2$ and CO rich, with
 eventual oxidation to CO$_2$ rate-limited by water photolysis and hydrogen escape. 
 
\end{abstract}


\newpage

\pagenumbering{arabic}


\section{Introduction}

The modern science of the origin of life on Earth begins with \citet{Haldane1929} and \citet{Oparin1938}. 
Both argued that a highly reduced early terrestrial environment --- profoundly unlike the world of today, even with O$_2$ removed --- 
was needed. 
Oparin's specific emphasis on methane, ammonia, formaldehyde, and hydrogen cyanide as primordial materials suitable for further development remains a recurring theme in origin of life studies \citep{Urey1952,Oro1961,Ferris1978,Stribling1987,Oro1990,Ricardo2004,Powner2009,Sutherland2016,Benner2019a}.
Although some of these materials --- formaldehyde in particular --- can be generated under weakly reducing conditions \citep{Pinto1980,Benner2019b}, others (such as cyanamide and cyanoacetylene) require strongly reducing conditions.         
The hypothesized reducing atmosphere inspired the famous and often-replicated Miller-Urey experiments, in which sparked or UV irradiated gas mixtures spontaneously generate a wide range of organic molecules \citep{Miller1953,Miller1955,Miller1959,Cleaves2008,Johnson2008}.

The geological argument against a reducing early atmosphere is nearly as old
 \citep[e.g.,][]{Poole1951}, 
although often accompanied by the caveat that things could have been different before the rock record
 \citep[e.g.,][]{Holland1964,Abelson1966,Walker1977,Holland1984}.
 The underlying presumption is that the atmosphere should have resembled volcanic gases.
 Modern volcanic gases are roughly consistent with the QFM (quartz-fayalite-magnetite) mineral buffer, for which the redox state is determined by chemical reactions between ferrous (Fe$^{+2}$) and ferric iron (Fe$^{+3}$).
  At typical magma temperatures, QFM predicts that H$_2$ and CO would be present at percent levels compared to H$_2$O
 and CO$_2$,
 and that methane and ammonia would be negligible. 
 
Some studies suggest that the Archean mantle had a similar redox state to today \citep{Delano2001,Canil2002,Rollinson2017},
while rare earth elements in zircons suggest a Hadean mantle consistent with QFM \citep{Trail2012}.
Large uncertainties in observationally derived oxygen fugacities ($\pm 2$ in $\log_{10}\!{\left(f_{\mathrm{O}_2}\right)}$) 
may obscure a secular trend, 
while some simplifying assumptions made in earlier $f_{\mathrm{O}_2}$ studies 
are open to question \citep{Wang2019}.
(The redox state of rocks is usually described by oxygen fugacity $f_{\mathrm{O}_2}$, 
which describes the formal abundance of O$_2$ gas in units of atmospheres.)
Two recent studies that use filtered samples hint that $\log_{10}\!{\left(f_{\mathrm{O}_2}\right)}$ of the mantle increased by
$\sim\!1.3$ from the early Archean to Proterozoic \citep{Aulbach2016,Nicklas2019}.

Concurrently, a body of experimental evidence has accumulated suggesting that ferrous silicates in Earth's mantle 
 disproportionate under great pressure into ferric iron and metallic iron, with the latter expected to migrate to the core \citep{Frost2008}.
This would leave the mantle, or at least part of the mantle, in a QFM-like state of oxidation from the time our planet first became big enough to be called Earth \citep{Armstrong2019}.

Given the incompatibility of a QFM mantle with a reduced atmosphere, workers have turned to impact degassing,
in which gases are directly released into the atmosphere on impact \citep{Matsui1986,Tyburczy1986}.
Most impactors are much more reduced than the mantle and often better endowed (gram per gram) in atmophile elements
\citep{Urey1952,Schaefer2007,Hashimoto2007,Sugita2009}.
Many meteorites, including ordinary chondrites and enstatite chondrites,
contain substantial amounts of metallic iron and iron sulfides.
Gases that equilibrate with these highly reduced meteoritic materials would be highly reduced themselves 
 \citep{Kasting1990,Schaefer2007,Hashimoto2007,Schaefer2010,Kuwahara2015,Schaefer2017},
provided that there is enough iron to reduce all the atmophiles in the impactor.  
But if there are more atmophiles to reduce than iron to reduce them, the gas composition can evolve to a much more oxidized state \citep{Schaefer2017}. 

Several of the new impact-degassing studies \citep{Hashimoto2007,Schaefer2007,Schaefer2010,Schaefer2017} 
calculate gas compositions in equilibrium with mineral assemblages
 at fixed pressures, with temperature treated as an independent variable.
These calculations often promise big yields of CH$_4$ and NH$_3$ at low temperatures.
However, actual yields depend on the quench conditions in the gas as it cools after the impact.
A cooling gas is said to have quenched when the chemical reactions maintaining equilibrium between species
become so sluggish that the composition of the gas freezes \citep{Zeldovich1967}.
Quenching is mostly determined by temperature.
Gas phase reactions for making CH$_4$ from CO are strongly inhibited by low temperatures, and
those for making NH$_3$ from N$_2$ are even more strongly inhibited, so unless
an abundant catalyst were available to lower the effective quench temperature \citep{Kress2004},
there is a tendency for the shock-heated gas to quench to CO, N$_2$, and H$_2$.
Of the new studies, only \citet{Kuwahara2015} have attempted to calculate quench conditions, but their results
are problematic because they used the entropy of shocked silica to 
estimate the entropy of shocked carbonaceous chondrites, which results in artificially low temperatures and artificially
large amounts of methane. 
Finally, in a full account, the quenched plume of impact gases would be mixed into,
and diluted by, the pre-existing atmosphere. 

\medskip

This study follows the lead of \citet{Genda2017a,Genda2017b} and \citet{Benner2019a} in addressing how the largest cosmic impacts
changed the ocean and atmosphere that were already present on Earth.
We go beyond \citet{Genda2017a,Genda2017b} and \citet{Benner2019a} in addressing not just the single largest impact but also a full range of sizes,
extending to impacts 100 and even 1000 times smaller (Hadean Earth would have experienced scores of these).
Our particular focus is on impacts that process the entire atmosphere and hydrosphere.
This differs from previous studies of smaller impacts that find that the main product of impact is HCN,
and this only in atmospheres with C/O ratios greater than unity \citep[cf.,][]{Chyba1992,Fegley1998}. 
Section 2 provides a brief summary of impacts after the Moon formed
as constrained by geochemistry and the craters of the Moon.
Section 3 addresses impact-generation of methane-rich atmospheres on Earth.
The emphasis is on impacts that are big enough to vaporize the oceans, as these
produce long-lasting hot conditions at high pressures, and thus can be highly favorable to methane and sometimes even to ammonia.
Section 4 uses a simple model to address the subsequent photochemical evolution of these atmospheres.
The emphasis here is on the fate of methane and the production of organic material and hazes,
 on the photochemistry of nitrogen and the generation of HCN and other nitrogenated organics, and on hydrogen escape.
Ammonia is (mostly) deferred to the Discussion.

\section{The Late Veneer}
\label{section:two}

The highly siderophile elements (HSEs) comprise seven heavy metals (Ru, Rh, Pd, Os, Ir, Pt, and Au) with
 very strong tendencies to partition into planetary cores.
If Earth's mantle and core were fully equilibrated almost all of its HSEs would be in the core, and the
tiny remnant in the mantle would be highly chemically fractionated \citep{Walker2009,Day2016,Rubie2015,Rubie2016}.
 But this is not what is seen.
Rather, the mantle contains a modest cohort of excess HSEs that, to first approximation,
 are present in roughly the same relative abundances that they have in chondritic meteorites \citep{Day2016}.  
One explanation is that the excess HSEs were dropped into the mantle and left stranded there some time after
 core formation was complete.
If the mantle's HSEs were added with other elements in chondritic proportions, they correspond to about 0.5\% of Earth's mass \citep{Anders1989}.
The late-added mass carrying the HSEs is usually called the ``late veneer.''  

 The late veneer measured in this way is very big.    
Viewed literally, 0.5\% of Earth's mass corresponds to a veneer 20 km thick.
 Gathered into a sphere, it corresponds to a rocky world 2300 km diameter
 --- as big as Pluto, and more massive.
We will call this the ``maximum HSE'' veneer.
If the veneer were sourced from fragments of differentiated worlds, the veneer mass could be a little 
smaller or much bigger.

Historically, the late veneer was presumed volatile-rich, as would be expected 
if the last materials to fall to Earth fell from the cold distant outer solar system
\citep{Anders1977,Wanke1988,Dreibus1989,Albarede2013}.
However, the late veneer now appears constrained
by Ru isotopes to resemble either enstatite chondrites, enstatite achondrites (aka aubrites), or iron meteorites of type IAB, 
and thus appears to come from the same deep inner solar system reservoir as Earth itself
 \citep{Dauphas2017,Fischer-Godde2017,Bermingham2018,Hopp2018b}.
 All of our samples of these materials are profoundly reduced and very dry.
 This apparently excludes the late veneer as the source of water on Earth \citep{Fischer-Godde2017},
and thus the late veneer can be presumed to have impacted into an Earth already fully plenished with oceans,
a view also consistent with oxygen isotopes \citep{Greenwood2018}.
The late veneer's role changes from water bearer to water changer: it must now be viewed as a source of reducing power injected
into Earth's near-surface environment \citep{Genda2017a,Genda2017b,Benner2019a}. 
    
The total reducing power delivered by the maximum late veneer can be illustrated by using all of its metallic iron
 to reduce water to hydrogen, in stoichiometry
 $\mathrm{Fe} + \mathrm{H}_2\mathrm{O}  \rightarrow \mathrm{FeO}  + \mathrm{H}_2$.
 The iron that accompanied the mantle's excess HSEs corresponded to $1\times 10^{25}$ g of metal.
 Because the HSEs remained in the mantle neither they nor the iron that came with them went to the core, and thus we can be
 confident that the iron was oxidized at the surface, in the crust, or in the mantle.
 There is enough iron in the late veneer to reduce $1.8\times 10^{23}$ moles of H$_2$O to H$_2$ and FeO, which corresponds
 to reducing of 2.3 oceans of water to hydrogen.  
   
 If the late veneer were characterized by size-number statistics typical of stray solar system bodies,
it is likely that most of the mantle's HSE excess was carried by a single Pluto-sized body \citep{Sleep1989,Tremaine1993,Bottke2010,Brasser2016,Genda2017a}.
Comparison with the uncertain but apparently much smaller lunar HSE excess \citep{Day2016}
is consistent with the conjecture that the maximum HSE event was singular \citep{Brasser2016,Morbidelli2018},
although this is not required, as there are other ways of explaining the scarcity of lunar HSEs that 
do not imply different accretion rates for Earth and Moon \citep[cf.,][]{Kraus2015}.   

 But even if the late veneer were delivered by one body,
  it does not follow that its mass was added to Earth in a moment.
 There is a considerable likelihood, estimated as 50\% by \citet{Agnor2004}, that an impact results not in a merger
 but rather in the disintegration of the smaller body.
 The debris are distributed in a ring around the Sun coincident with Earth's orbit
 and swept up by Earth over tens or hundreds of thousands of years \citep{Genda2017a,Genda2017b}.
Few of the debris are swept up by the Moon, owing to the much greater gravitational cross-section of Earth
 with respect to debris in quasi-circular orbits \citep{Genda2017a}.
 This kind of distributed event is likely to strand nearly all of its HSEs in the crust or at shallow depths in Earth's mantle, while
  the direct impact of a Pluto-sized body might be expected to drive much of the impactor's core directly into our own.
  Stranding {\em all} the newly added HSEs in the mantle without fractionation fulfills a second independent requirement imposed by the mantle's Ru isotopes, which were not mass-fractionated by partitioning between the mantle and core \citep{Fischer-Godde2017,Hopp2018b}.
Dividing the impactor into myriads of smaller particles would also be more effective 
 at chemically reducing Earth's atmosphere and ocean. 
 For example, \citet{Genda2017b} model a Moon-sized impactor and find that 60\% of
 the iron would be divided into mm-size droplets.
 The overall picture resembles that suggested by \citet{Urey1952}, who wrote that
``materials would have fallen through the atmosphere in the form of iron and silicate rains and would have reacted with the atmosphere [and hydrosphere] in the process.''

An important caveat is that a maximum HSE impact may not couple well to the oceans.
A Pluto-sized impact would blanket Earth in tens of kilometers of impact ejecta, which 
 is so much deeper than the oceans that much of the iron may have been buried before it could react with water.
Under these conditions, the buried iron would have remained unoxidized in the upper mantle for a considerable period of time.
We know from the presence of the HSEs and the unfractionated Ru isotopes that the iron was not removed to the core.
The iron must therefore have strongly influenced the redox state
of volcanic gases until its oxidation was complete.  The effect of this
is to prolong the influence of the maximum HSE event to geological time scales.

 \begin{table}[tb]  
 \small
\caption{Some Representative Hadean impacts }
\begin{tabular}{l r r r r r r r r r r r }
 \multicolumn{6}{l}{ \large\strut} &  \multicolumn{5}{l}{Products, dry atmosphere$^a$ [bars]} \\
 Category & N$^b$ & $M_i^c$ [g] & Vap$^d$ & Red$^e$ & CO$_2^f$        & H$_2$  &CO & CO$_2$ & CH$_4$ & NH$_3$  \\ 
  \hline
Max HSE & 0-1 & 2(25)   & 200 & 2 & 100             & 57 & 5(-6) & 1.4(-5) & 9.0 & 0.08 \\
%
~~~~QFI$^g$  &   &     &   &   & 100                          &  74  & 1(-5) & 2(-6) & 7.6 & 0.050 \\
~~~~IW$^g$  &   &     &   &   & 100                          &  35  & 4(-5) & 6(-4) & 13.7 & 0.086 \\
~~~~QFM$^g$ &   &     &   &   & 100                          &  10  & 0.11 & 65.6 & 13.6 & 0.01 \\
 Pretty Big$^h$ &  0-2  & 2.5(24)  & 20 &  0.2  & 20         & 7.6 & 5(-4) & 0.06 & 2.9 & 0.03 \\
                  &        &          &        &        & 5           & 7.4 & 6(-6) & 4(-4)& 0.34 & 0.01 \\
``Ceres'' &  1-4 & 1(24)   & 8 &  0.08 & 5              & 3.9 & 3(-4) & 0.06 & 0.52 & 0.006  \\
  ``Vesta''  & 2-10  & 2.5(23)  & 2 & 0.02 & 5             & 3.9 & 0.06 & 1.6 & 0.17 & 0.002 \\
                &        &       &    &         & 2             & 2.6 & 6(-4) & 0.4 & 0.054 & 0.0015 \\
                &        &       &    &         & 1             & 2.0 & 2(-4) & 0.14 & 0.023 & 0.0011  \\
 ~~~~QFM$^i$  &    &    &    &       & 2             & 1.8 & 1(-5) & 0.02 & 0.28 & 0.06 \\
Sub-Vesta & 3-20  & 1(23)  & 0.8 & 0.008      & 5       & 2.7 & 0.005 & 2.7 & 0.008 & 8(-4) \\   
%
 ~~~~QFM$^i$ &   &    &    &    & 2                    & 1.5  & 6(-5)  & 0.1  & 0.36 & 0.037  \\  
%
%
 S.Pole-Ait.$^j$ & 10-100  & 1(22) & 0.1 & 8(-4)  & 2       & 0.37 & 0.008 & 1.5 & 1(-7) & 2(-5) \\  
     ~~~~QFM$^i$ &   &    &    &    & 2                        & 0.65 & 0.015 & 1.32 & 1(-6) & 7(-5) \\  %
\hline
\multicolumn{11}{l}{$a$ -- {\Large\strut}The dry atmosphere presumes that all water has condensed at the surface.}\\
\multicolumn{11}{l}{$b$ -- Number of Hadean impacts in each class, bracketed between minimum and maximum veneer}\\
\multicolumn{11}{l}{$c$ -- $M_i$ presumes 33\% metallic iron, like EH (high iron enstatite) chondrites or bulk Earth}\\
\multicolumn{11}{l}{$d$ -- Oceans of water that could be vaporized by the impact}\\
\multicolumn{11}{l}{$e$ -- Potential reducing power of the impact, expressed as oceans of water that can be reduced to H$_2$}\\
\multicolumn{11}{l}{$f$ -- Atmospheric CO$_2$ before the impact [100 bars = 2300 moles cm$^{-2}$]}\\
 \multicolumn{11}{l}{$g$ -- Assumed to equilibrate with the named mineral buffer (defined in Appendix A)}\\
 \multicolumn{11}{l}{$h$ -- Plausible size of biggest impact in a minimum late veneer}\\
 \multicolumn{11}{l}{$i$ -- Atmosphere and ocean are assumed to equilibrate with QFM buffer at 650 K}\\
 \multicolumn{11}{l}{$j$ -- South Pole-Aitken is the largest impact basin preserved on the Moon}\\
  \end{tabular}
\label{table_two}
\end{table}%

\medskip
 There is also a small chance that the late veneer is an illusion.
 It has been suggested that Earth's HSE excess may date to the Moon-forming impact itself \citep{Newsom1989,Sleep2016,Brasser2016}.
  If so, the mantle's HSE excess overestimates the amount of reducing power delivered to Earth after the Moon-forming impact.
 A crude lower bound on the late veneer  
 can be extrapolated from the lunar crater record \citep{Sleep1989,Zahnle1997,Zahnle2006}.
 This scaling suggests that the ``minimum late veneer'' delivered between 3-30\% of the mass as the maximum HSE veneer.
 The large uncertainty, and large total mass striking Earth compared to the Moon, both arise from the high
 probability that {\em all} the largest bodies in a given population hit the Earth \citep{Sleep1989}.  
 
 Table \ref{table_two} lists a representative sampling of maximum and minimum late veneer impacts.
 The number of bodies in any given size class is estimated from the cumulative relation $N(>\!m)\propto m^{-b}$,
 with $0.5\!<\!b\!<\!0.9$, with the smaller number based on the number of lunar basins
 and with the larger number based on the total cumulative mass of the maximum
 late veneer using methods described by \citet{Zahnle1997}.
 The number of oceans that can be vaporized assumes that 50\% of the impact energy is available to evaporate
 an ocean ($1.4\times 10^{24}$g) of water and heat the steam to 1500 K.
 The number of oceans that can be reduced to H$_2$ assumes an EH enstatite chondritic composition with 33\% Fe
 by mass and that the reaction
  $\mathrm{Fe}+\mathrm{H}_2\mathrm{O} \rightarrow \mathrm{FeO}+ \mathrm{H}_2$ goes to completion.
   Other entries in Table \ref{table_two} are discussed as they arise.

 Evidence has recently emerged that Earth's molybdenum --- another siderophile element, but somewhat less so than the HSEs ---
 has an isotopic composition distinct from Earth's HSEs \citep{Budde2019}. 
 This has been interpreted by its discoverers to mean that Theia --- the name widely given to the Moon-forming impactor --- was made of different stuff than the late veneer \citep{Budde2019}.
 \citet{Budde2019} even suggest that Theia was the source of Earth's water, although in our opinion
 it seems equally plausible that Earth's distinctive Mo predates the Moon-forming impact.
 From our perspective here it makes little difference whether Earth's water was delivered by Theia or predated Theia,
  because in either case the water was present on Earth when the late veneer came.


\section{Thermochemical Model}
\label{section: Thermochemical Model}

The redox state of gases in equilibrium with rocks is often described by mineral buffers
that govern the capacity of the rock to consume or release oxygen.
Three such buffers are described in Appendix A.
Mineral buffering is most likely to matter when the rock-to-atmophile ratio is large, as it is
in meteorites or for Earth-like planets considered as a whole.
Mineral buffers are less obviously appropriate for describing the interaction of meteorites with oceans
and atmospheres that are much bigger than the meteorite \citep{Elkins-Tanton2008a}.
Only the very biggest post-Moon-forming impacts are big enough for a mineral buffer
set by the impactor to apply on a global scale. 
For anything smaller, the oxygen in the atmosphere and ocean much exceeds the reducing power
in the impactor, and hence the reduced mineral buffers are exhausted before the atmosphere and ocean can
fully equilibrate \citep{Elkins-Tanton2008a,Schaefer2017}.
What this means is that, in most cases, a better approximation than hewing to a mineral buffer
is to stoichiometrically remove the oxygen scavenged by metallic iron from the atmosphere and ocean,
and then compute the resulting equilibria amongst the gases.

\subsection{Equilibrium chemistry}
\label{section:Equilibrium chemistry}

We solve for five potentially major gases --- H$_2$, H$_2$O, CO, CO$_2$, and CH$_4$ --- while presuming that other gases are minor.
In particular, we treat nitrogen as a minor perturbation, and we ignore sulfur and chlorine.
We treat the equilibrium chemistry of the atmosphere as a whole.
We solve for the column number densities $N_j$ and for the partial pressures $p_j$ of the 5 species.
The total pressure $p$ is the weight of the atmosphere,  
\begin{equation}\label{total pressure}
 p = g\sum_j N_j m_j ,
 \end{equation}
 where $g$ is the gravity and $m_j$ is the mass of species $j$.
Partial pressures $p_j$ are related to column densities and the total pressure by
\begin{equation}\label{partial pressure}
 p_j = {p N_j \over \sum_j N_j}.
 \end{equation}
Note that, in general, $p_j \neq N_j m_j g$; i.e., partial pressures are proportional to number fractions, not mass fractions. 

In the five gas system, hydrogen and carbon are conserved:
\begin{equation}
\label{hydrogen conservation}
N_{\mathrm{H}} = 2N_{\mathrm{H}_2} + 2N_{\mathrm{H}_2\mathrm{O}} + 4N_{\mathrm{CH}_4}
\end{equation}
and
\begin{equation}
\label{carbon conservation}
N_{\mathrm{C}} = N_{\mathrm{CO}} + N_{\mathrm{CO}_2} + N_{\mathrm{CH}_4} .
\end{equation}
In the absence of a mineral buffer, oxygen is also conserved,
\begin{equation}
\label{oxygen conservation}
N_{\mathrm{O}} = N_{\mathrm{H}_2\mathrm{O}} + N_{\mathrm{CO}} + 2N_{\mathrm{CO}_2} .
\end{equation}
The other two relations needed to close the system are chemical equilibria.  We use the water gas shift reaction
\begin{equation}
\label{shift}
\refstepcounter{reaction}\tag{R\arabic{reaction}}
  {\rm CO} + {\rm H}_2{\rm O} \leftrightarrow {\rm CO}_2 + {\rm H}_2 ,
\end{equation}
which has equilibrium constant
\begin{equation}
\label{shift reaction}
K_{\ref{shift}} = \frac{p_{\mathrm{H}_2\mathrm{O}}\, p_{\mathrm{CO}} } {p_{\mathrm{H}_2}\, p_{\mathrm{CO}_2}} 
\approx 18.28\exp{\left( -2375.6/T - 5.69\times 10^5/T \right)} ,
\end{equation}
and the corresponding reaction for methane,
\begin{equation}
\label{methane}
\refstepcounter{reaction}\tag{R\arabic{reaction}}
  {\rm CO} + 3{\rm H}_2 \leftrightarrow {\rm CH}_4 + {\rm H}_2{\rm O} ,
\end{equation}
which has equilibrium constant 
\begin{equation}
\label{methane reaction}
K_{\ref{methane}} = \frac{p_{\mathrm{CH}_4}\, p_{\mathrm{H}_2\mathrm{O}} } { p_{\mathrm{CO}}\, p^3_{\mathrm{H}_2} } 
\approx 5.239\times 10^{-14} \exp{\left( 27285/T \right)} \;\;\mathrm{atm}^{-2} .
\end{equation}
As is customary, partial pressures in Equations \ref{shift reaction} and \ref{methane reaction} are in atmospheres.
Equilibrium constants given here are low order curve fits \citep{Zahnle2014} generated using thermochemical data from \citet{Chase1998}. 

When oxygen is controlled by a mineral buffer, 
oxygen is not conserved and a third chemical equilibrium reaction is needed to link the system to the mineral buffer.
The mineral buffer supplies the oxygen fugacity $f_{\mathrm{O}_2}$, which has units of pressure.
We use
\begin{equation}
\label{oxygen}\refstepcounter{reaction}\tag{R\arabic{reaction}}
  2{\rm H}_2{\rm O} \leftrightarrow {\rm O}_2 + 2{\rm H}_2 ,
\end{equation}
with equilibrium constant
\begin{equation}
\label{water reaction}
K_{\ref{oxygen}} = \frac{p^2_{\mathrm{H}_2\mathrm{O}} } {  p^2_{\mathrm{H}_2}\, f_{\mathrm{O}_2}} 
\approx 1.158\times 10^{-6} \exp{\left( 59911/T \right)} \;\;\mathrm{atm}^{-1} .
\end{equation}
We will suppose that the gas remains equilibrated with the mineral buffer until the metallic iron 
is either exhausted or physically removed from interaction with the gas. 
This fixes the total oxygen content of the atmosphere.
Thereafter the gas phase chemistry continues to evolve with oxygen conserved in response to further cooling
until the gas phase reactions themselves quench.

\medskip
It is convenient to treat nitrogen species as minor perturbations, solved separately for fixed amounts of the five
important CHO species.
Separating N also facilitates taking into account that nitrogen species quench at higher temperatures than H, C, and O.   
This simplification is accurate provided that NH$_3$ is not a major gas.
Nitrogen is conserved,
\begin{equation}
\label{nitrogen conservation}
N_{\mathrm{N}} = N_{\mathrm{NH}_3} + N_{\mathrm{HCN} } + 2N_{\mathrm{N}_2} .
\end{equation}
Two chemical equilibria are needed, one for ammonia
 \begin{equation}
\label{ammonia}\refstepcounter{reaction}\tag{R\arabic{reaction}}
  {\rm N}_2 + 3{\rm H}_2 \leftrightarrow 2{\rm NH}_3 ,
 \end{equation}
with equilibrium constant
\begin{equation}
\label{ammonia reaction}
K_{\ref{ammonia}} = \frac{p^2_{\mathrm{NH}_3}} {p_{\mathrm{N}_2}\, p^3_{\mathrm{H}_2}} 
\approx 5.90\times 10^{-13} \exp{\left( 13207/T \right)} \;\;\mathrm{atm}^{-2} ,
\end{equation}
and another for HCN,
 \begin{equation}
\label{HCN}\refstepcounter{reaction}\tag{R\arabic{reaction}}
  {\rm H}_2{\rm O} + {\rm HCN} \leftrightarrow {\rm CO} + {\rm NH}_3 ,
 \end{equation}
with equilibrium constant
\begin{equation}
\label{cynanide reaction}
K_{\ref{HCN}} = \frac{ p_{\mathrm{H}_2\mathrm{O}}\, p_{\mathrm{HCN}} } {p_{\mathrm{CO}}\, p_{\mathrm{NH}_3} }
\approx 1.99\exp{\left( -6339.4/T \right)} .
\end{equation}
Equations \ref{ammonia reaction} and \ref{cynanide reaction} reduce to a quadratic equation for NH$_3$.
In practice HCN is never produced abundantly by shock heating in the large impacts and H$_2$O-CO$_2$ atmospheres considered in this study.
The chief source of HCN in this study is photochemical, discussed in Section 4 below.

\subsection{Quenching}
\label{section:quenching}

Chemical reactions are generally fast at high temperatures and chemical equilibria are quickly established between major species.  
As the gas cools, chemical reactions between the more stable molecules 
slow down until for all practical purposes they stop and the gas composition is said to have quenched or frozen \citep{Zeldovich1967}.
Here we ignore possible catalysts and address only gas phase chemistry, which is the most pessimistic case for methane and ammonia.
We employ two quench points, one for the H$_2$-H$_2$O-CH$_4$-CO-CO$_2$ system
and a significantly hotter one for the H$_2$-N$_2$-NH$_3$-HCN system.

We have characterized quench conditions 
for CO hydrogenation to CH$_4$ and for N$_2$ hydrogenation to NH$_3$ in brown dwarf atmospheres \citet{Zahnle2014}.
There we devised curve fits to global quench temperatures for the key chemical systems using
a time-stepping thermochemical kinetics code employing nearly 100 chemical species and more than 1000 chemical reactions.
Our curve fits are degenerate between total pressure and the H$_2$ partial pressure, because these are the same in brown dwarfs.
For making CH$_4$, our model predicts that quenching is linear with $p$ at low pressures but quadratic with $p$ at high pressure.
The low pressure quench temperature is
\[
T_{q1}\left(\mathrm{CH}_4\right) = \frac{42000\,\mathrm{K}}{\ln{\left(3.3\times 10^5 t_c p\right)}} .
\]
The timescale $t_c$ is in seconds.
The high pressure form is
\[
T_{q2}\left(\mathrm{CH}_4\right) = \frac{25000\,\mathrm{K}}{\ln{\left(0.025\,t_c p^2\right)}} .
\]
The quench temperature is the smaller of the two,
\begin{equation}
\label{CH4_quench}
T_{q}\left(\mathrm{CH}_4\right) = \min{\left( T_{q1}\left(\mathrm{CH}_4\right), T_{q2}\left(\mathrm{CH}_4\right)\right)}.
\end{equation}
Other published estimates of quench conditions in the CH$_4$-CO-H$_2$ system   
\citep[c.f.,][]{Prinn1977,Visscher2011,Line2011} are similar enough
that they also predict CH$_4$-dominated atmospheres for the cases where we predict them;
the different chemical quenching times are explicitly compared in \citet{Zahnle2014}.   

\begin{figure}[!tbh]
 \centering
  \includegraphics[width=0.9\textwidth]{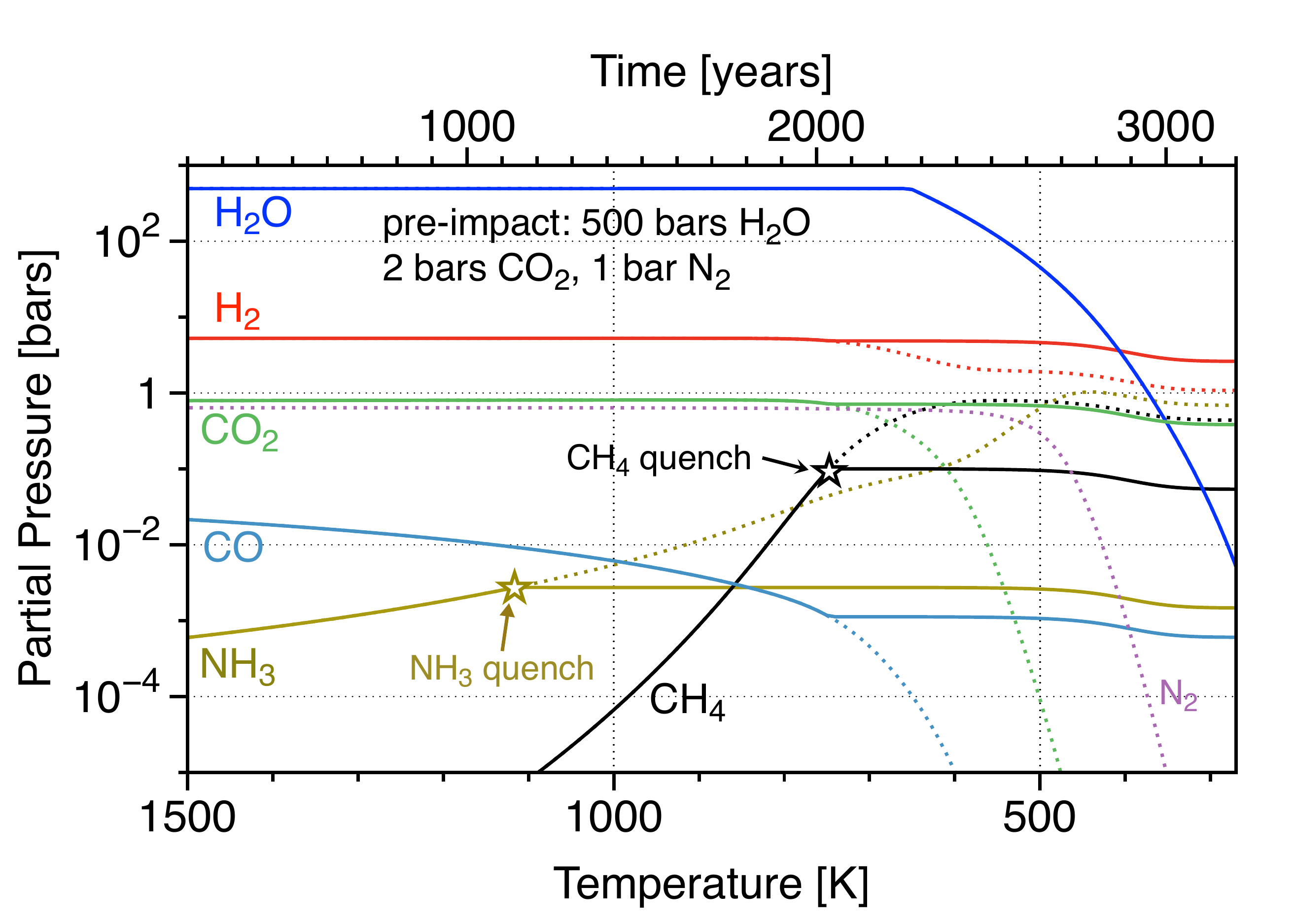}  
\caption{Example of quenching after a Vesta-sized impact into an Earth with
a pre-impact atmosphere containing 2 bars CO$_2$, 1 bar of N$_2$, and 1.85 oceans (500 bars) of liquid water.
Cooling (time) marches monotonically from left to right.
Quench points for CH$_4$ and NH$_3$ H$_2$-N$_2$-NH$_3$ are indicated with stars.
Solid lines show quenched compositions while dotted lines extend the equilibria to temperatures colder than the quench point
for gas phase reactions.
Partial pressures are also affected by water condensation ($T<650$ K).}
 \label{Quenching_Vesta}
\end{figure}

\medskip
Quenching in the NH$_3$-N$_2$ system occurs at higher temperatures,
\begin{equation}
\label{NH3_quench}
T_{q}\left(\mathrm{NH}_3\right) = \frac{52000\,\mathrm{K}}{\ln{\left(1.0\times 10^7 t_c p\right)}} .
\end{equation}
Typically $T_{q}\left(\mathrm{NH}_3\right)$ is about 300 K warmer than $T_{q}\left(\mathrm{CH}_4\right)$.
There is considerable uncertainty regarding the mechanisms of N$_2$ hydrogenation, with resulting considerable
uncertainty in quenching times.  The older estimate by \citet{Prinn1981} predicts slower chemistry,
while two more recent estimates \citep{Line2011,Zahnle2014} give similar results for conditions encountered here.
In practice the different kinetics predict similar chemical compositions \citep{Zahnle2014},
because the NH$_3$/N$_2$ ratio is not strongly sensitive to temperature.  
Figure \ref{Quenching_Vesta} illustrates quenching after a Vesta-scale impact.

\subsection{Cooling times}
\label{section:cooling}

Impacts that vaporize the oceans create globally hot, high pressure conditions that can last for thousands of years.
The energy invested in evaporating water and heating the major atmospheric gases in an ocean-vaporizing impact is
\begin{equation}
\label{E_atm}
E_{atm} = \left(Q_w + C_{\mathrm{H}_2\mathrm{O}} {\Delta T} \right) M_{\mathrm{H}_2\mathrm{O}} + C_{\mathrm{CO}_2} {\Delta T} M_{\mathrm{CO}_2} + C_{\mathrm{N}_2} {\Delta T} M_{\mathrm{N}_2}
\end{equation}
where $C_{\mathrm{H}_2\mathrm{O}} = 2\times 10^7$ ergs/g/K, $C_{\mathrm{CO}_2} = 8\times 10^{6}$ ergs/g/K,
and $C_{\mathrm{N}_2} = 1.1\times 10^{7}$ ergs g$^{-1}$ K$^{-1}$ are heat capacities of H$_2$O, CO$_2$, and N$_2$, respectively; $Q_w=2.5\times 10^{10}$ ergs g$^{-1}$ is the latent heat of vaporization of H$_2$O at 273 K; and
${\Delta T} \sim 1500$ K approximates heating the atmosphere to a point where rock vapors become significant.
Evaluated for relevant parameters,
\begin{equation}
\label{E_atm evaluated}
E_{atm}[\mathrm{ergs}] = 8\times 10^{34} \left(M_{\mathrm{H}_2\mathrm{O}} \over 1.4\times 10^{24}\mathrm{~g} \right) +
 6\times 10^{33} \left(M_{\mathrm{CO}_2} \over 5\times 10^{23}  \mathrm{~g} \right) + 8\times 10^{31} \left(M_{\mathrm{N}_2} \over 5\times 10^{21}  \mathrm{~g}\right)
\end{equation}
where the fiducial masses correspond to an ocean of water, a 100 bar CO$_2$ atmosphere, and a 1 bar N$_2$ atmosphere, respectively.

Evaporating the oceans and heating the steam to the temperature of the condensing rock vapor are the big terms in the energy
budget for impacts of this scale.  
This energy is compared to the impact energy
\begin{equation}
\label{E_i}
E_i[\mathrm{ergs}] = {1\over 2} m_i v_i^2 
= 1.5\times 10^{35} \left( m_i \over 10^{23} \mathrm{~g} \right)  \left( v_i \over 17 \mathrm{~km}\mathrm{s}^{-1}  \right)^2 .
\end{equation}
If half of the impact energy is spent heating and vaporizing water 
(with the other half deeply buried and unavailable on timescales shorter than thousands of years,
or promptly radiated to space at higher temperatures in the immediate aftermath of the event),
a Vesta-size impact can evaporate and heat 2.3 oceans of water.
The maximum HSE impact exceeds the Vesta-size impact by a factor of 100.
Put another way, an EH-like (high iron, enstatite chondritic) impact creates 100 times more steam than hydrogen.

The characteristic cooling time $t_c$ is approximated by how long it takes for the steam atmosphere to cool to the quench point.
Quench temperatures for methane and ammonia will be hotter than water's critical point, so we ignore
the latent heat released by condensation. 
\begin{equation}
\label{t_c}
t_c = \frac{ C_{\mathrm{H}_2\mathrm{O}} \left(T-T_q\right)  M_{\mathrm{H}_2\mathrm{O}} + C_{\mathrm{CO}_2} \left(T-T_q\right) M_{\mathrm{CO}_2} + C_{\mathrm{N}_2} \left(T-T_q\right) M_{\mathrm{N}_2}} { A_{\oplus}F_{ir}} ,
\end{equation}
where $A_{\oplus}$ is the area of the Earth and $F_{ir}\approx 1.5\times 10^5$ ergs cm$^{-2}$s$^{-1}$ is the radiative cooling rate of
a terrestrial steam atmosphere with a 50\% albedo after the Sun has reached the main sequence (ca.\ 50 Myr).
This cooling rate is valid provided that water clouds condense somewhere in the atmosphere \citep{Abe1988,Nakajima1992}.
 For methane, for which quench temperatures are of order 800 K, the
relevant cooling is from 1400 K to 800 K.  Evaluated,
\begin{equation}
\label{t_c evaluated}
t_c [\mathrm{s}] =  2\times 10^{10} \left(M_{\mathrm{H}_2\mathrm{O}} \over 1.4\times 10^{24}  \mathrm{~g} \right) 
+ 3\times 10^9\left(M_{\mathrm{CO}_2} \over 5\times 10^{23}  \mathrm{~g} \right) +  4\times 10^7\left(M_{\mathrm{N}_2} \over 5\times 10^{21}  \mathrm{~g} \right) ,
\end{equation}
which is of the order of 1000 years for most cases we consider. 
For NH$_3$, whose quench temperature is $\sim$ 300 K hotter than methane's, the cooling time is about half as long.

If an impact is too small to fully vaporize the ocean, the ocean remains cool and acts as a heat sink
that competes with thermal radiation to space.
In the relevant case, the impact leaves the atmosphere much hotter than the ocean and mostly made of H$_2$ and water vapor.
The lower atmosphere will therefore be stable against convection.
Under these conditions the flow of energy down to the ocean is limited by radiative transfer.
To illustrate, compare the diffusive flux of downward radiation in the Eddington approximation (any astronomy textbook),
\begin{equation}
\label{F_down}
F_{\downarrow} = {16\over 3}{\sigma_B T^3\over \kappa \rho}{dT\over dz} ,
\end{equation} 
to Earth's net cooling rate $F_{ir}$.
The gray approximation opacity of water vapor is $\kappa \approx 0.1$ cm$^2$g$^{-1}$ \citep{Nakajima1992}.
Assume 30 bars of 1100 K steam as an example,
for which the density at the surface is $\rho=0.025$ g cm$^{-3}$ and the surface temperature is 500 K (both set by the boiling point).
The temperature gradient appropriate to cooling the whole atmosphere is $600\, (=\!\!1100-500)$ K over a 20 km scale height.
With ${dT/dz} = 3\times 10^{-4}$ K cm$^{-1}$, we estimate that $F_{\downarrow} \approx 0.5\times 10^5$ ergs cm$^{-2}$s$^{-1}$,
which is 30\% of net cooling ($F_{ir}$) to space.  
The radiative flux $F_{\downarrow}$ is relatively small because $\rho\kappa$ is big.
 (In this example, $F_{\downarrow}$ would exceed $F_{ir}$ for impacts that generate less than 10 bars of steam.) 
 We conclude that, in general, relatively little of the energy
in a hot deep steam atmosphere will flow downward to the ocean.
The chief exception would be if there is enough CO$_2$ that the much hotter atmosphere
is nonetheless dense enough to sink in cooler steam.
This might happen for smaller impacts in deep CO$_2$ atmospheres, and if it did, it
would result in complications that we will not address here.

\subsection{Results}
\label{section:results}

We consider three classes of impact.

(i) If the impact is big enough, it delivers enough iron to fully reduce all the H$_2$O and CO$_2$ at the surface. 
Under these conditions, the $\mathrm{Fe} + \mathrm{H}_2\mathrm{O} \leftrightarrow \mathrm{FeO}+ \mathrm{H}_2$ equilibrium 
is likely to govern the oxidation state of the atmosphere throughout the cooling phase.  
Earth's maximum HSE impact was in this size range. 
The iron-w{\"u}stite (IW) buffer describes the simple reaction of iron and steam to make FeO (w{\"u}stite) and H$_2$,
and so is likely to be kinetically favored in the short term.
On longer timescales the more reducing quartz-fayalite-iron buffer (QFI, effectively a buffer between iron and olivine) may dominate.
Both mineral buffers favor CH$_4$,
but the QFI buffer would also consume most of the H$_2$O in favor of H$_2$.
The latter outcome resembles the story proposed to explain the desiccation of Mars by \citet{Dreibus1989,Kuramoto1997}. 

A caveat is that, for impacts of this scale, the global ejecta blanket should have been tens of kilometers thick.
Although the molten iron in the ejecta must have passed through the atmosphere and ocean to reach the surface
--- \citet{Genda2017b} estimate that more than half the iron is initially disseminated as mm-sized droplets ---
it is plausible that some of the iron was buried before it could react with H$_2$O or CO$_2$.
If so, some of the reducing power of the impact would at first have been sequestered from the surface,
 only becoming available as reduced gases emitted to the atmosphere on a longer geologic timescale.
 Modeling the fate of iron in the post-impact mantle is beyond the scope of this study.

(ii) The oceans are fully vaporized but there isn't enough metallic iron in the impact to reduce all of the H$_2$O in the ocean to H$_2$.
Impacts on this scale leave much of the H$_2$O and CO$_2$ unreacted.
In such an impact the ejecta blanket is not much thicker than the ocean is deep, and
conditions at the surface are supercritical for water, promoting efficient chemical coupling of the water with the iron while the iron lasts
\citep[see][and references therein]{Choudhry2014}. 
There are of the order of ten such impacts in a representative maximum late veneer, and 1-3 in a minimum late veneer.
We assume that in these events the introduced Fe consumes oxygen from the water and CO$_2$  
until the all the injected Fe is gone.  Thereafter the atmosphere evolves with its oxygen content (in H$_2$O, CO, and CO$_2$)
held constant.
In these events the steam atmosphere is deep, thick, and hot, and cooling is slow, conditions that
strongly favor CH$_4$ and, to a lesser degree, NH$_3$. 
 
 (iii) The impact is too small to fully vaporize the oceans.  These events feature faster cooling times 
 and lower atmospheric pressures, with the amount of steam generated proportional to the energy released by the impact.
 The lower pressures are generally much less favorable to CH$_4$ formation, 
but small impacts are interesting because there are more of them and they are likelier to be survived by life or its precursors.
For small impacts, we will find that the QFM mineral buffer often generates a more reduced gas composition than predicted from
 scavenging by impact iron of the oxygen in the hydrosphere and atmosphere. 
  For these events we will presume that the ferrous iron already present in the
crust is available as an additional sink of oxygen at the QFM buffer.
These matters are discussed in more detail below.

\medskip
We treat the volume of the ocean and the amount and state of carbon in the atmosphere before the impact as initial
conditions.  For water, we assume that 1.85 oceans of water (5 km) were present at the surface.
Bigger oceans allow for more extensive loss of hydrogen to space without desiccating the planet.
We take the view, provisionally, that a much drier planet ($\ll$1 ocean) will not evolve to Earth as we know it.

\begin{figure}[!tbh]
 \centering
  \includegraphics[width=0.9\textwidth]{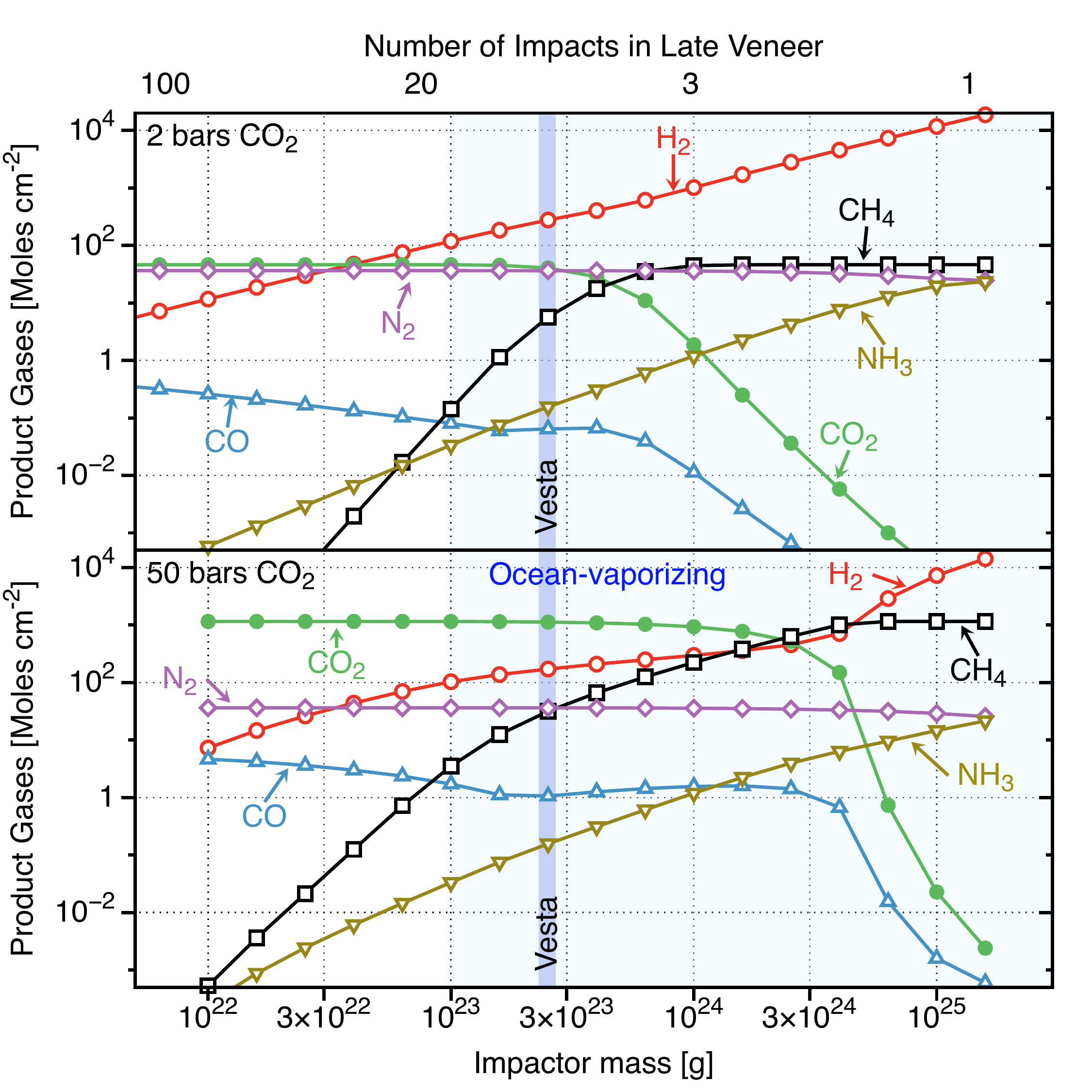}   
\caption{Quenched impact-induced transient atmospheres as a function of impactor mass (EH composition) for two pre-impact
atmospheres, one with 50 bars of CO$_2$ and the other with 2 bars.
Both have 1 bar of N$_2$ and 1.85 oceans (500 bars) of water on the surface.
Inventories are shown in moles to highlight the chemical transformation.
A Vesta-size impact is indicated by the shaded vertical bar.  
Potentially ocean-vaporizing impacts are indicated by lighter shading.
A rough guide to the number of impacts of a given size is listed across the top.
In these models, all reducing power is furnished by the quantitive reduction of the impactor's Fe to FeO; no mineral buffering is assumed.
In the maximum late veneer impact (at right), nearly all CO and CO$_2$ are converted to CH$_4$.}
 \label{Products_vs_Mi}
\end{figure}

Carbon reservoirs are not well constrained.  Between surface, crust, and mantle, Earth may
hold the equivalent of $200\pm 100$ bars of CO$_2$ \citep{Sleep2001a,Dauphas2014a}.
One end-member is hot and oxidized, with CO$_2$ being initially divided roughly equally between
 a melted QFM mantle and Henry Law partitioning of 100 bars of CO$_2$ in the atmosphere in the aftermath
 of the Moon-forming impact
  \citep{Holland1984,Abe1997,Zahnle2007,Elkins-Tanton2008b}. 
An oxidized mantle could have been consequent to a previous history of hydrogen escape or
to iron-mineral disproportionation \citep{Frost2008}.  
CO$_2$ can also be generated from thermal decomposition of carbonate minerals if these were near the surface.
Although we do not explicitly consider more reduced atmospheres (CO or CH$_4$) as initial conditions,
we will find below that thick CO or CH$_4$ atmospheres can be long-lasting in the Hadean.
The CO$_2$ atmosphere is the most oxidized and hence the most conservative case.
We treat $p_{\mathrm{CO}_2}$ as a free parameter.

\begin{figure}[!tbh]
 \centering
  \includegraphics[width=0.9\textwidth]{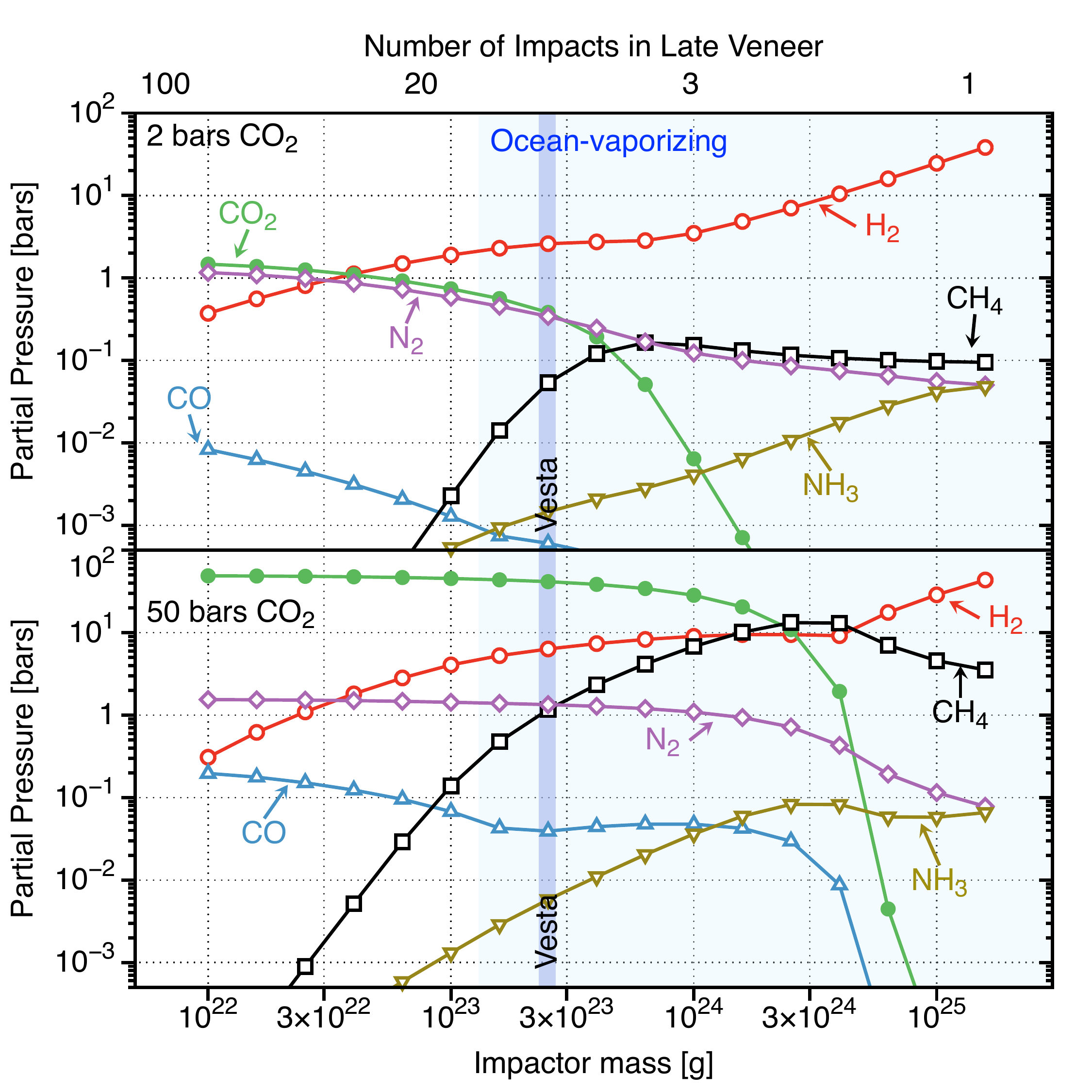}   
\caption{Same as Figure \ref{Products_vs_Mi}, but atmospheric compositions shown as partial pressures.}
 \label{Pressure_vs_Mi}
\end{figure}

\begin{figure}[!tbh]
 \centering
  \includegraphics[width=0.9\textwidth]{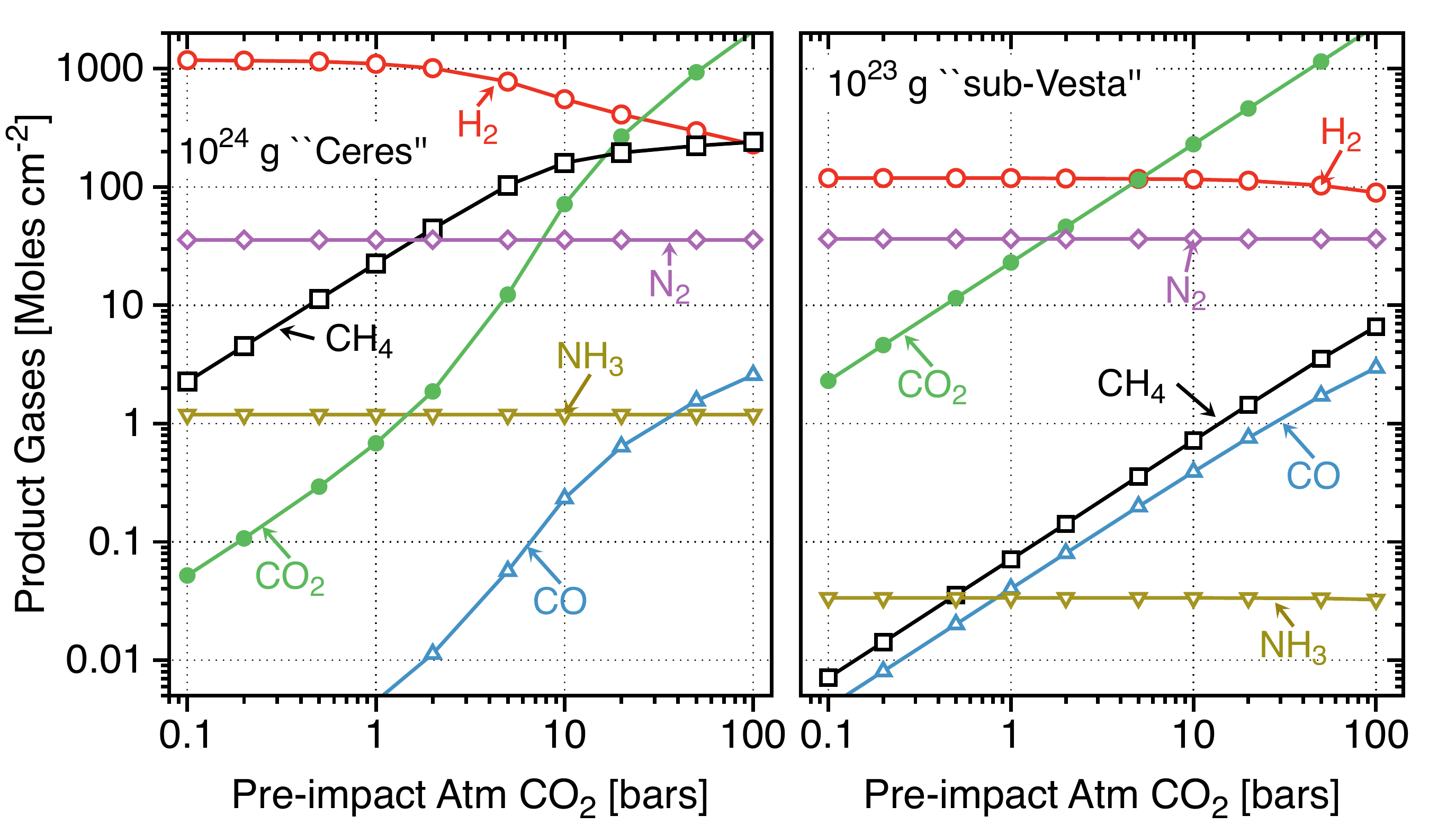}  
\caption{Quenched impact-induced transient atmospheres as a function of how much CO$_2$ was in the atmosphere before
the impact, for two sizes of impact, one ocean-vaporizing, one not.
The pre-impact Earth has 1 bar of N$_2$ and 1.85 oceans (500 bars) of water on the surface.
The Ceres-sized ($10^{24}$ g) impact (EH) is big enough to convert thinner CO$_2$ atmospheres
to CH$_4$, but hasn't enough Fe to fully reduce the thicker CO$_2$ atmospheres. 
The smaller ``sub-Vesta'' impact is too small to fully vaporize 1.85 oceans; 
the chief effect is to make a lot of H$_2$.
}
 \label{Products_vs_CO2}
\end{figure}

Table \ref{table_two} lists a sampling of possible Hadean impacts.
Before impact, Earth is presumed to have had 1.85 oceans of water at the surface (500 bars, 28.3 kmols cm$^{-2}$) and
 one bar (36 moles cm$^{-2}$) of N$_2$ in the atmosphere.
 The amount of CO$_2$ varies between examples.
 ``100 bars'' of CO$_2$ corresponds to 2300 moles cm$^{-2}$. 

Figures \ref{Products_vs_Mi} and \ref{Pressure_vs_Mi} show post-impact atmospheres
for a wide range of impact sizes for 2 and 50 bars of CO$_2$.
Figure \ref{Products_vs_Mi} shows column densities (moles cm$^{-2}$), which makes the chemical trends clear,
while Figure \ref{Pressure_vs_Mi} shows the same information as partial pressures.  
 The reducing power of the impactors presumes high iron EH (enstatite) or H (ordinary) chondritic bodies (33\% Fe by mass, which also approximates the bulk Earth). 
Impact energy assumes an impact velocity of 17 km s$^{-1}$.
Table \ref{table_two} lists major product gases for each case.
Many of the cases listed in the table are used as initial conditions for photochemical evolution in Section 3 below. 
If the Fe is incompletely used up, the corresponding impact mass can simply be scaled up; i.e.,
if half the Fe goes unreacted, the required impactor would have twice the mass.

The maximum HSE impact delivers marginally enough Fe to fully reduce the atmosphere and hydrosphere,
which suggests that equilibration with a mineral buffer may be plausible.
Table \ref{table_two} lists several maximum HSE cases with 100 bars of CO$_2$ equilibrated to different buffers.
The QFI buffer, which would also fully reduce the atmosphere and hydrosphere, seems
likeliest if the reactions all go to completion.
But if much of the iron is buried before it reacts, a more oxidized buffer might be more reasonable.

 In the next category, the ``Pretty Big'' cases approximate the biggest impact in our lower bound late veneer,
whilst ``Ceres'' and ``Vesta'' are impacts with the mass of the real Ceres and the real Vesta.
These are all ocean-vaporizers, but none deliver nearly enough iron to fully reduce the ocean.
Figure \ref{Products_vs_CO2} compares outcomes
as a function of $p$CO$_2$ for Ceres-sized impacts.

The third category is represented in Table 1 by two ``sub-Vestas'' and the lunar South Pole-Aitken impact
that do not fully evaporate the oceans.
These are small enough that life or its precursors might survive. 
These impacts are also too small to deliver enough metallic iron
to reduce the ocean to the QFM composition.
This means that the reducing power of Earth's mineral buffers --- made active by the heat of the impact ---
needs to be taken into account.
The South Pole-Aitken basin is an example of a relatively minor event.

\begin{figure}[!tbh]
 \centering
  \includegraphics[width=0.9\textwidth]{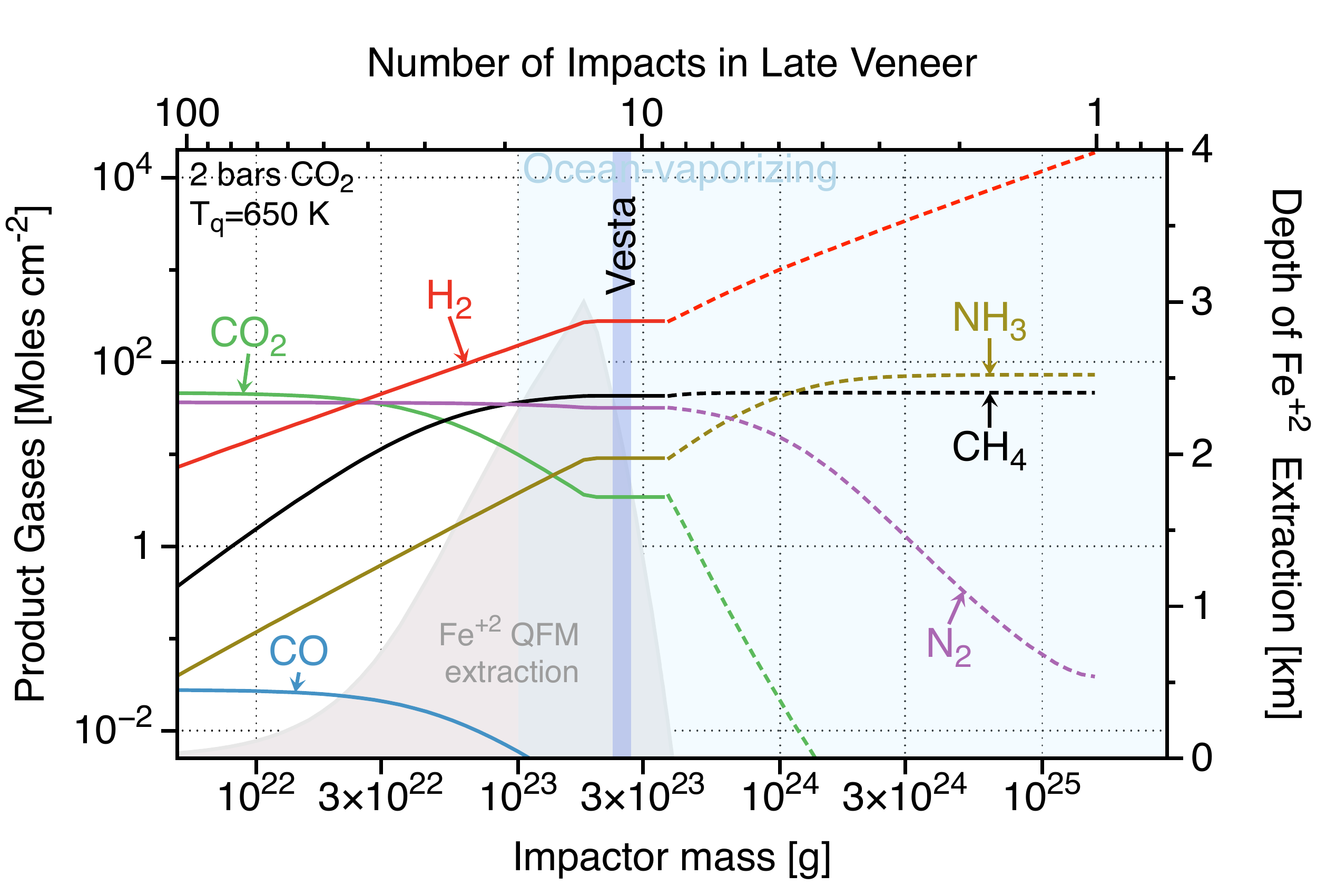}   
\caption{The left-hand axis shows gas compositions as a function of impact mass when quenching is assumed to take place at 650 K (water's critical temperature).
Solid lines for smaller impacts ($M_i<4\times 10^{23}$ g) show gas compositions in equilibrium with the QFM mineral buffer.
The right-hand axis shows the depth (shading) to which FeO in the crust must be oxidized to magnetite (Fe$_3$O$_4$)
to maintain equilibrium with the QFM buffer at 650 K. 
Dashed lines for gas compositions generated by bigger impacts ($M_i>4\times 10^{23}$ g) are determined by metallic iron delivered by the impact. 
These latter compositions are no more reduced than the otherwise comparable compositions seen for gas phase quenching in Figure \ref{Products_vs_Mi} above,
but the assumed low 650 K quench temperature favors CH$_4$ and NH$_3$ over H$_2$.
}
 \label{QFM_crust_C2_T650}
\end{figure}

Table \ref{table_two} lists two sub-Vestas.
The first uses only the reducing power of the impact. 
This is the most pessimistic case.
The other assumes equilibration with a crustal QFM mineral buffer, but at an arbitrary lower quench temperature of 650 K.  
In effect, the second case asks what happens if supercritical water was in itself enough to
ensure that the coupled H$_2$-H$_2$O-CO-CO$_2$-CH$_4$-N$_2$-NH$_3$ system 
equilibrated on a thousand year time scale.
This can be viewed as the optimistic limit on what sub-ocean-vaporizing impacts can do to generate species like CH$_4$ and NH$_3$.

Figure \ref{QFM_crust_C2_T650} illustrates the potential inherent in the more optimistic case.
Here we assume that the atmosphere and ocean remain chemically equilibrated with the crust or mantle at the
 QFM buffer while water remains supercritical; i.e., we set $T_q=650$ K.
The QFM buffer is only weakly reducing because it has only Fe$^{+2}$ to offer as a reductant.
On the other hand there is a great deal of Fe$^{+2}$ available in 
impact-heated crust and mantle materials,
although the crustal source is not inexhaustible.

To illustrate these considerations, in Figure \ref{QFM_crust_C2_T650} we estimate the depth in the crust (global average) to 
which FeO must be oxidized to Fe$_3$O$_4$ (magnetite), assuming
that the crust was 10\% FeO by mass and that all the FeO is oxidized to Fe$_3$O$_4$, and
taking into account the reducing power delivered by the impact as metallic iron.
Figure \ref{QFM_crust_C2_T650} presumes a pre-existing atmosphere with the equivalent of 2 bars of CO$_2$ and one bar of N$_2$.
The figure shows that, even in a Vesta-scale impact,
the required reducing power can be extracted from FeO in the uppermost 3 km of a QFM crust.
Smaller impacts use less of Earth's FeO because they don't evaporate the entire ocean, while  
larger impacts (larger than $4\times 10^{23}$ g in Figure \ref{QFM_crust_C2_T650}) 
deliver more metallic iron than needed to maintain QFM.

\section{Photochemical evolution of impact-generated transient atmospheres}
\label{section:Photochemical evolution}

Our goal in this Section is to model the photochemical decay of the impact-generated transient reduced atmosphere.
 In particular we are interested in what happens to methane.
  The key processes driving the atmosphere's evolution are ultraviolet photolysis and hydrogen escape.
  Thus, fundamentally, we are most concerned with counting the photons and apportioning their effects.
  
\subsection{The Photochemical Model}
\label{section:Photochemical Model}

In the photochemical model, we consider six major species: H$_2$, CH$_4$, H$_2$O, CO$_2$, CO, and N$_2$.
Minor species include HCN (nitriles), C$_2$H$_n$ (a mix of C$_2$H$_2$, C$_2$H$_4$, C$_2$H$_6$),
and organic haze.
Other molecules and free radicals that are considered include
NO, NH, N($^4$S), N($^2$D), O($^3$P), O($^1$D), $^3$CH$_2$, $^1$CH$_2$, CH$_3$, and OH.
We refer to ground state N($^4$S) as N, ground state O($^3$P) as O, and ground state $^3$CH$_2$ as CH$_2$.
Key reactions are listed in Appendix B.
Atomic H is implicit and lumped with H$_2$ for accounting purposes. 
Ions are not explicitly included, although the first order effects of ion chemistry are taken into account as loss processes for CO$_2$ and CH$_4$
(Appendix C).

Our purpose is to construct the simplest model that captures the first order consequences of photochemical evolution
of unfamiliar atmospheres,
 while conserving elements and counting the photons.
  Anything more complicated (e.g., a 1-D atmospheric photochemistry code) would necessarily
 introduce several poorly-constrained free parameters.  
We therefore assume that the major atmospheric constituents are uniformly distributed vertically,
  affected only by the totality of chemical and physical sources and sinks.  
   We treat each major species $j$ as a column density $N_j$ (units of number per cm$^2$).
  Ultraviolet photons are sorted into several spectral windows and  
  the effects of photolysis are apportioned in accordance with the first order consequences of photochemistry;
  these simplifications will be discussed in detail below.
  The columns are evolved through time by integrating $dN_j/dt$.
  Hydrogen escape and the effectively irreversible photolysis of methane impose direction.    
      
   \medskip
  There are two first order complications to the simplest model that demand attention. 
  First, H$_2$O --- usually the most abundant gas after the impact --- condenses to make
  oceans.  Thereafter its abundance at stratospheric altitudes where photolysis
  takes place is limited by the atmosphere's cold trap.
  Water vapor is key to these models because water vapor is often the major oxidant.  The drier the stratosphere, the
  more slowly it evolves and the more likely it is to favor reduced products like hydrocarbons.
  A next generation study might investigate the water vapor contents of self-consistent radiative-convective atmospheres,
  but this level of modeling goes beyond the scope of this study. 
  Here we treat stratospheric water vapor in the atmosphere
  as a free parameter. 
  
  \medskip
  The second complication is the shadow cast by organic hazes. 
  Organic hazes are expected when methane is abundant and subject to UV photolysis \citep{Trainer2006,Horst2012,Horst2018b}.
  The analogy is Titan.
   We expect Earth's hazes, when present, to be optically thick, much thicker than on Titan today,
  because UV irradiation of early Earth was at least 1000 times greater than of modern Titan.
   Here we follow \citet{Wolf2010} and parameterize haze optical depth as a function of haze production rate. 
 Because water photolysis would be the major source of oxidants in a methane-rich atmosphere,
  the competition between hazes and water for UV photons creates a positive feedback in which one or the other
  dominates.    
  
\subsubsection{Irradiation}
\label{section:Irradiation}

We divide the solar FUV and EUV spectrum into six windows that align with particular properties
of the atmosphere (Table \ref{tab:EUV}). 
We neglect the Lyman continuum (80-91.2 nm, current photon flux of $8\times 10^9$ cm$^{-2}$s$^{-1}$) as filtered out by atomic H.
Absorption at wavelengths longer than the Lyman continuum usually leads to dissociation of molecules into
two neutral species. 
Absorption at wavelengths shorter than the Lyman continuum usually ionizes the molecule, leaving the molecule
provisionally intact as an ion.    

The ancient Sun was a stronger source of EUV radiation than is the modern Sun.
We scale the different channels according to the general rule that hotter source regions are relatively more enhanced
by solar activity and were therefore relatively more enhanced
when the Sun was young \citep{Zahnle1982,Claire2012}.  
Enhancement factors for the different windows are listed in Table \ref{tab:EUV}.
    
\begin{table}[tbh]
\small
   \centering
   \topcaption{Fluxes [cm$^{-2}$s$^{-1}$] and cross sections [cm$^2$]}
   \begin{tabular}{@{} llrrrrrr @{}} 
   \toprule
 & \multicolumn{7}{c}{Spectral Window} \\
  & & EUV$_1^a$ & EUV$_2^b$ & EUV-N$_2^c$ & Lyman $\alpha$ & FUV-CO$_2$ &  FUV-H$_2$O \\ 
       \midrule
  & $i$ & 1 & 2  & 3 & 4  & 5 & 6   \\
 & $\lambda$  [nm] & $<80$ & $91.2\!-\!117$ & $91\!-\!100$ & $121.5$  & $125\!-\!170$ & $170\!-\!185$   \\
  & $F_{\lambda}^d$ & $3\!\times\! 10^{10}$ & $2\!\times\! 10^{10}$ & $3.6\!\times\! 10^9$ & $3.6\!\times\! 10^{11}$ & $4\!\times\! 10^{11}$ &  $2.5\!\times\! 10^{12}$  \\
  & $S_{\lambda}^e$ & $30$ & $20$ & $10$ & $10$ & $7$ & $4$  \\
   \midrule
 $j$ & & \multicolumn{6}{c}{Cross section $\sigma_{ij}$ [cm$^2$]} \\
  1 &H$_2$ & $2\!\times\! 10^{-18}$ & $2\!\times\! 10^{-18}$  & --   & -- & -- & --  \\
 2&CH$_4$ & $2\!\times\! 10^{-17}$ & $8\!\times\!10^{-18}$  & $8\!\times\! 10^{-18}$   & $8\!\times\! 10^{-18}$  & -- & -- \\
 3&H$_2$O & $1.3\!\times\! 10^{-17}$ & $8\!\times\! 10^{-18}$ & $8\!\times\! 10^{-18}$ & $8\!\times\! 10^{-18}$ & $1\!\times\! 10^{-18}$ & $3\!\times\! 10^{-18}$ \\
 4&CO$_2$ & $2\!\times\! 10^{-17}$ & $4\!\times\! 10^{-17}$  & $4\!\times\!10^{-17}$  &  $5\!\times\! 10^{-20}$  & $8\!\times\! 10^{-19}$ & --\\
 5&CO &  $1.3\!\times\! 10^{-17}$ &  -- &  -- & --& --& --\\
 6&N$_2$ & $1.3\!\times\! 10^{-17}$ &  -- &  $2.5\!\times\! 10^{-16}$  & --& --& --\\
   \bottomrule
\multicolumn{8}{l}{$a$ -- Photo-ionizing EUV excluding the Lyman continuum}\\
\multicolumn{8}{l}{$b$ -- EUV$_2$ excludes photoionizing radiation and radiation that photolyzes N$_2$}\\
\multicolumn{8}{l}{$c$ -- EUV-N$_2$ is the portion of non-photoionizing EUV that coincides with N$_2$ absorption} \\
\multicolumn{8}{l}{$d$ -- Quiet Sun irradiance, photons cm$^{-2}$s$^{-1}$ at 1 AU} \\
\multicolumn{8}{l}{$e$ -- Young Sun enhancement over modern quiet Sun}\\
  \end{tabular}
   \label{tab:EUV}
\end{table}

Total column photolysis rates $\Phi_j$ [cm$^{-2}$s$^{-1}$] for species $j$ allow for competition for photons between species,
\begin{equation}
\label{total photolysis}
\Phi_j = \frac{1}{4}  \sum_{i=2,6} F_iS_i {\sigma_{ij}N_j \strut\displaystyle  \over \strut\displaystyle \sum_{k=1,7} \sigma_{ik} N_k } ,
\end{equation}
where $k$ is an index running over the species.
This approach conserves photons, which is the principal requirement here.
Photo-ionization is treated separately,
\begin{equation}
\label{photoionization}
\Phi^{\ast}_j = \frac{1}{4} F_1S_1 {\sigma_{1j}N_j  \over \sum_k \sigma_{1k} N_k } .
\end{equation}

 Organic hazes when present may provide UV protection to the deeper atmosphere \citep{Sagan1997,Pavlov2001,Wolf2010}.
\citet{Wolf2010} constructed a microphysical model of organic hazes of early Earth.
They considered spherical haze particles and ``fractal'' haze particles, with the latter model much preferred by its authors.
We fit power laws to the fractal haze optical depths listed in their Table S1  
as a function of the haze production rate,
\begin{equation}
\label{Wolf_uv}
 \tau_{uv} \approx 10 \left( dN_{\mathrm{haze}}/dt \over 3\times 10^{10} \right)^{0.8}
 \end{equation}
 and
\begin{equation}
\label{Wolf_vis}
\tau_{vis} \approx 0.5 \left( dN_{\mathrm{haze}}/dt \over 3\times 10^{10}\right)^{0.7} .
 \end{equation}
The ultraviolet optical depth refers to 197 nm and   
the visible optical depth refers to 564 nm.
We have converted units from production in grams per year to the equivalent number
of carbon atoms cm$^{-2}$s$^{-1}$, which are the units used in this paper.
Haze production rates on early Earth can exceed $1\times 10^{12}$ carbon atoms cm$^{-2}$s$^{-1}$,
which corresponds to UV and visible fractal haze optical depths of the order of 160 and 6, respectively.  

Hazes can suppress H$_2$O photolysis if the stratosphere is dry.  
This can lead to a positive feedback that encourages haze formation.
As hazes thicken, there is less H$_2$O photolysis and less oxidation,
which favors more haze formation.
 The consequence of this positive feedback resembles a phase change, 
 in which much of the carbon derived from methane photolysis polymerizes into
 a wide range of heavier, generally oxygen-deficient organics (which we loosely refer to as ``haze'') that precipitate to the 
 troposphere and probably to the surface.   
 
 In wetter stratospheres, haze formation competes with oxidation consequent to H$_2$O photolysis.
 The organics and hazes that form under these conditions will contain more oxygen, suggesting a relatively greater role for acids,
 aldehydes, carbonyls, and other more water-soluble molecules that are more likely to rain out when they reach the troposphere.
 These may be essential ingredients for genesis of ribose for the RNA world \citep[cf.,][]{Benner2019b}.
 

\subsubsection{Photolysis}
\label{section:Photolysis}

Methane photolysis is dominated by Lyman $\alpha$ radiation.
Methane photolysis at Lyman $\alpha$ mostly yields 
an excited methylene radical $^1$CH$_2$ plus hydrogen \citep{Huebner1992}. 
Singlet methylene can be collisionally de-excited to the less reactive triplet $^3$CH$_2$,
or it can react with CH$_4$ or H$_2$ to make CH$_3$ radicals.
We will refer to these small radicals generically as CH$_n$.
Both CH$_2$ and CH$_3$ react very quickly with atomic N from N$_2$ photolysis to make C-N bonds
or with atomic O from CO$_2$ photolysis to make C-O bonds.
The C-O bonds once formed are difficult to break photochemically. 
The N reactions are the primary sources of HCN in a CH$_4$-N$_2$ irradiated atmosphere.
We neglect the possible catalytic role of N$_2$ through the CH$_2$N$_2$ (diazomethane) intermediary.

Competition for Ly $\alpha$ photons is limited.
We do not expect scattering by atomic hydrogen to be significant, because the solar Ly $\alpha$ emission is
much broader than the velocity dispersion in hydrogen atoms at atmospheric temperatures.  
CO$_2$ has a very small cross section to Ly $\alpha$, only about 0.5\% of methane's.
Both H$_2$ and CO have hot absorption lines that partially overlap with Lyman $\alpha$ emission.
Resulting fluorescence has been seen in planetary nebulae \citep{Lupu2006} and cometary comae \citep{Lupu2007}, respectively. 
However, the effect requires ro-vibrationally excited H$_2$ and CO molecules, and hence is unlikely 
to be important at relevant conditions.
Water's cross section at Ly $\alpha$ is about the same as methane's, but we do not
expect H$_2$O to be abundant at the highest altitudes once the cold trap has been established in the lower atmosphere.  
What this all means is that, when methane is abundant, Ly $\alpha$ photochemistry takes place high in
the atmosphere, aligning it with N$_2$ photolysis and generally favoring the production of organic
hazes and nitriles.
When methane is not abundant, Ly $\alpha$ photochemistry takes place deeper in the atmosphere,
which better aligns methane photolysis with CO$_2$ photolysis and H$_2$O photolysis,
an alignment that favors methane oxidation and disfavors nitrile production.

Nitrogen (N$_2$) photolysis is dominated by several very strong narrow absorption bands that coincide with solar emission lines.
\citet{Huebner1992} stresses the importance of the overlap between a strong N$_2$ band 
and solar Ly $\gamma$ (92.25 nm).
Comparison between the solar spectrum \citep{Curdt2001} and the N$_2$ absorption spectrum predicted by
\citet{Liang2007}, \citet{Li2013}, and \citet{Heays2014} 
shows that strong N$_2$ absorption bands coincide with several other Lyman lines, including 
Ly $\delta$ (94.974 nm), Ly $\epsilon$ (93.78 nm), Ly 7, 8, and 12, and also with an NIII line (99.17 nm).  
There is also overlap at 91.3 nm where Lyman lines ($n\!>\!30$) pile up in the approach to the Lyman limit. 
\citet{Huebner1992} estimated from a suite of discordant experiments
that the N$_2$ photolysis cross section at Ly $\gamma$ is of order $2.5 \times 10^{-16}$ cm$^2$.

Nitrogen's chief competition for photons is with CO$_2$, which has big cross sections of order $4\times 10^{-17}$ cm$^2$
at wavelengths where N$_2$ absorbs.
Molecular hydrogen also absorbs at some of these wavelengths, Ly $\gamma$ in particular,
 but cross sections are generally smaller than $10^{-18}$ cm$^{2}$. 
 For the quiet Sun at Earth, we estimate a total N$_2$ photolysis
rate of $3.6\times 10^9$ cm$^{-2}$s$^{-1}$,
which agrees well with what \citet[][their Figure 2]{Liang2007} compute for Titan when scaled to 1 AU.
 
Photolysis between 91.2 nm and 100 nm splits N$_2$ into a ground state N atom and an electronically excited
$\mathrm{N}(^2\mathrm{D})$.
The $\mathrm{N}(^2\mathrm{D})$ can react with H$_2$ or CH$_4$ to make NH, react with CH$_4$ to make CH$_2$NH, and react with CO$_2$ to make CO and NO \citep{Herron1999}. 
Both NH and NO react rapidly with N to reconstitute N$_2$.
Otherwise the most important N and NH reactions are  
with hydrocarbon radicals to make HCN 
and other nitriles, such as acetonitrile (CH$_3$CN) and cyanoacetylene (HCCCN). 
Photolytic production of HCN has been predicted to work well in an N$_2$-CO$_2$ atmosphere \citep{Zahnle1986a,Tian2011}.
Photochemical organics have been hypothesized as a source of reduced nitrogen and reduced carbon that can be subducted by the mantle \citep{Wordsworth2016}.  

At the top of the atmosphere,
CO$_2$ photolysis is dominated by EUV wavelengths between 91.2 nm (the Lyman limit) and 115 nm.
Weaker absorption at FUV wavelengths between 130 nm and 180 nm can be as important if other absorbers
are not abundant.
At FUV wavelengths, CO$_2$ photolysis usually creates a ground state CO molecule and a highly reactive O($^1$D) atom.
\begin{equation}
\label{O singlet D}
\mathrm{CO}_2 + h\nu \rightarrow \mathrm{CO} + \mathrm{O}(^1\mathrm{D})
\end{equation}
At the shorter wavelengths, photolysis can also yield electronically excited CO and a ground state O atom \citep{Huebner1992}.
We will assume that excited CO is de-excited by collisions.

The $\mathrm{O}(^1\mathrm{D})$ atom is highly reactive, including reaction with CH$_4$ to liberate CH$_3$.   
Key $\mathrm{O}(^1\mathrm{D})$ reactions are listed in Appendix B.
 It can be de-excited to the less reactive $\mathrm{O}(^3\mathrm{P})$ ground state
by collisions with CO$_2$, N$_2$, and CO, but it reacts quickly with CH$_4$ and H$_2$.
By contrast, reactions of ground state O with CH$_4$ and H$_2$ are negligibly slow at 300 K.
We therefore take the initial reaction of $\mathrm{O}(^1\mathrm{D})$ with CH$_4$ as the rate-limiting step for CH$_4$ loss from CO$_2$ photolysis.
Subsequent reactions of O with free radicals like CH$_3$ are fast and result in CO bonds.

Carbon monoxide has a similar spectrum to N$_2$, but unlike N$_2$, few of its bands align with strong
solar emission lines.  
CO dissociation into neutral atoms is dominated by the Lyman continuum \citep{Huebner1992}, and hence
is relatively unimportant in hydrogen-rich atmospheres.

Water photolysis yields OH and H for FUV with $\lambda < 190$ nm. 
If the stratosphere is very dry, organic hazes have potential to shield H$_2$O from photolysis, especially at 
wavelengths $\lambda > 182$ nm, where H$_2$O's cross section starts to fall off rapidly with increasing $\lambda$.  

Because we do not distinguish between H and H$_2$, hydrogen photolysis is important only as opacity. 

Ammonia is swiftly photolyzed by UV radiation between $185<\lambda<215$ nm at wavelengths  
 where H$_2$O and CO$_2$ absorptions are very weak.
 The products are highly reactive NH and NH$_2$ radicals \citep{Huebner1992}.
 These can lead to N$_2$ formation, but they can also react with hydrocarbons if the latter are plentiful.

\subsubsection{The methane budget}
\label{section:methane}

Methane ends up either as organics or HCN, or is oxidized to CO or CO$_2$.
While CH$_4$ is preponderant, photochemistry following photolysis will tend to generate hydrocarbons \citep{Lasaga1971,Yung1978, Zahnle1986a,Trainer2006,Horst2018b} and nitrogenous organics.
When CO$_2$ is preponderant, methane is more often oxidized to formaldehyde (HCHO) or CO.

Methane can also be oxidatively attacked by products of H$_2$O photolysis and CO$_2$ photolysis.
The most important of these are O($^1$D) atoms from CO$_2$ photolysis (R32) and OH from H$_2$O photolysis
or reaction of O($^1$D) with H$_2$. 
(Reactions of CH$_4$ with H and ground state O atoms are slow unless the gas is much hotter than we have assumed, while
reaction with N($^2$D) typically creates nitriles.) 
OH reacts fairly rapidly with CO to make CO$_2$, and more slowly at room temperature
with H$_2$ or CH$_4$ to yield H$_2$O.
The reactions of OH with H$_2$ and CH$_4$ are sensitive to temperature, whilst the reaction with CO is not.
At 300 K, $\mathrm{OH}+\mathrm{CO}\rightarrow \mathrm{CO}_2$ is about $20\times$ faster than reaction with H$_2$ or CH$_4$,
and at 250 K its about $100\times$ faster.

Column oxidative loss of CH$_4$ is approximated 
by loss reactions with O($^1$D) from CO$_2$ photolysis and OH from H$_2$O photolysis:
\begin{equation}
\label{dCH4 ox}
\left({d N_{\mathrm{CH}_4} \over dt} \right)_{\!\!ox} = -\Phi_{\mathrm{CO}_2} \left(  {k_{22}N_{\mathrm{CH}_4} \over \sum_j k_{2j} N_j } \right)
-\Phi_{\mathrm{H}_2\mathrm{O}} \left(  {k_{32}N_{\mathrm{CH}_4} \over \sum_j k_{3j} N_j } \right) .
\end{equation}
The total CH$_4$ budget sums the photolytic, photoionic, and oxidative losses, 
\begin{equation}
\label{dCH4dt}
 {dN_{\mathrm{CH}_4} \over dt} = -\Phi_{\mathrm{CH}_4} + \left({d N_{\mathrm{CH}_4} \over dt} \right)_{\!\!ions} 
+ \left({d N_{\mathrm{CH}_4} \over dt} \right)_{\!\!ox}.  
\end{equation}

\subsubsection{Hydrocarbons and organic hazes}
\label{section:haze}

Reactions between CH, CH$_2$ and CH$_3$ lead to acetylene, ethylene, ethane, and eventually to more complicated hydrocarbons
that can form a high altitude haze.
The chief competing reactions are those with O or OH radicals.
%
We lump acetylene, ethylene, and ethane together as C$_2$H$_n$ hydrocarbons.
%
%
We equate the creation of CH$_n$ radicals to the
appropriate destruction rate of CH$_4$,
\begin{equation}
\label{dCHndt}
\frac{d N_{\mathrm{CH}_n}} {d t} = \Phi_{\mathrm{CH}_4} - \left(\frac{d N_{\mathrm{CH}_4}} {d t} \right)_{\!\!ions} - \left(\frac{d N_{\mathrm{CH}_4}} {d t} \right)_{\!\!ox}\geq 0 .
 \end{equation}
 We equate the production of oxidizing radicals in the haze-forming regions to the appropriate photolysis rates of H$_2$O and CO$_2$, 
 \begin{equation}
\label{dOxdt}
\frac{d N_{\mathrm{ox}}} {d t} 
= \Phi_{\mathrm{CO}_2}  + \Phi^{\!\ast}_{\mathrm{CO}_2}  + \Phi_{\mathrm{H}_2\mathrm{O}}  \geq 0 .
 \end{equation}
We assume that C$_2$H$_n$ molecules form when a CH$_n$ radical reacts with another CH$_n$ radical,
while CO forms when a CH$_n$ radical reacts with an O or OH radical,
\begin{equation}
\label{dC2Hndt}
\frac{d N_{\mathrm{C}_2\mathrm{H}_n}} {d t} 
= \frac{1}{2} \frac{d N_{\mathrm{CH}_x}} {d t} 
\left(  d N_{\mathrm{CH}_x}/d t  \over 
 d N_{\mathrm{ox}}/d t + d N_{\mathrm{CH}_x}/d t   \right)^{\!2} .
 \end{equation}
Organic hazes form when several CH$_n$ radicals react with the growing
polymer for each reaction with an O or OH,
\begin{equation}
\label{dhazedt}
\frac{d N_{\mathrm{haze}}} {d t} 
=  \frac{d N_{\mathrm{CH}_x}} {d t} 
\left(  d N_{\mathrm{CH}_x}/d t  \over 
 d N_{\mathrm{ox}}/d t + d N_{\mathrm{CH}_x}/d t  \right)^{\!m} .
 \end{equation}
For specificity we take $m\!=\!5$ (corresponding to six carbons).
For accounting purposes we assume that all hazes fall to the surface and accumulate without further reaction,
\begin{equation}
\label{tar}
N_{\mathrm{haze}} = \int \frac{d N_{\mathrm{haze}}} {d t} dt .
\end{equation}
Hazes are generally ineffective at shielding molecules from photolysis at wavelengths where an abundant gas absorbs strongly,
because hazes make up a very small mass fraction of the atmosphere. 
Where hazes can matter is in shielding a gas of very low abundance at wavelengths that would otherwise be transparent.
The cases of interest here are H$_2$O, which can be cold-trapped, and NH$_3$.  

\subsubsection{CO and CO$_2$ budgets}
\label{section:CO}

The CO$_2$ budget is a balance between photolytic losses and CO oxidation,
\begin{equation}
 \label{dCO2dt}
{dN_{\mathrm{CO}_2} \over dt} = 
-\Phi_{\mathrm{CO}_2} \left(1 - f_{2x} f_{35} \right)
-\Phi_{\mathrm{CO}_2}^{\ast}
+f_{35} \Phi_{\mathrm{H}_2\mathrm{O}}
+ \Phi_{\mathrm{CO}_2}\left(1- f_{2x}f_{45} \right) 
\end{equation}
where $f_{2x}$ denotes the fraction of O($^1$D) that react with other atmospheric species to create OH radicals,
\begin{equation}
 f_{2x} \equiv  {k_{21}N_{\mathrm{H}_2} + k_{22b} N_{\mathrm{CH}_4} +  2k_{23}N_{\mathrm{H}_2\mathrm{O}} \over \sum_j k_{2j} N_j } ,
\end{equation}
$f_{35}$ is the fraction of OH radicals that react with CO,
\begin{equation}
 f_{35} \equiv {k_{35}N_{\mathrm{CO}} \over \sum_j k_{3j} N_j },
\end{equation}
and $f_{45}$ is the fraction of ground state O atoms that react with CO to make CO$_2$,
\begin{equation}
 f_{45} \equiv {k_{45}N_{\mathrm{CO}} \over \sum_j k_{4j} N_j } .
\end{equation}
The latter reaction, although spin-forbidden, is important at high pressure in a dry CO-rich atmosphere in the absence of catalysts.

The CO budget reverses the CO$_2$ budget, and also includes the net oxidation of CH$_4$ as a source,
\begin{equation}
 \label{dCOdt}
{dN_{\mathrm{CO}} \over dt} = - {dN_{\mathrm{CO}_2} \over dt} - {dN_{\mathrm{CH}_4} \over dt} 
- {dN_{\mathrm{haze}} \over dt} - {dN_{\mathrm{HCN}} \over dt} ,
\end{equation}
while treating precipitation of organic hazes and nitriles as a carbon sink.

 This model of CO and CO$_2$ gives a better description of the sum of CO and CO$_2$ than it does of CO and CO$_2$ individually.
Within the model, speciation between CO and CO$_2$ is sensitive to H$_2$O (the only oxidant).  
We suspect that our model overpredicts CO at the expense of CO$_2$.

 \subsubsection{Nitrogen and HCN budgets}
\label{section:Nitrogen}

In the anoxic atmospheres relevant to this study,
nitrogen chemistry leads either to nitriles (e.g., HCN) or to the reconstitution of N$_2$.
The direct products of nitrogen photolysis are
\begin{equation}
{dN_{\mathrm{N}} \over dt} = \Phi_{\mathrm{N}_2}\left(1 + {k_{15} N_{\mathrm{CO}} +  k_{16}N_{\mathrm{N}_2} \over \sum_j k_{1j} N_j } \right),
\end{equation}
\begin{equation}
{dN_{\mathrm{NO}} \over dt} = \Phi_{\mathrm{N}_2}{k_{13} N_{\mathrm{H}_2\mathrm{O}} +  k_{14}N_{\mathrm{CO}_2} \over \sum_j k_{1j} N_j } ,
\end{equation}
\begin{equation}
{dN_{\mathrm{NH}} \over dt} = \Phi_{\mathrm{N}_2}{k_{11} N_{\mathrm{H}_2} +  k_{12b}N_{\mathrm{CH}_4} \over \sum_j k_{1j} N_j } ,
\end{equation}
and 
\begin{equation}
{dN_{\mathrm{H}_2\mathrm{CN}} \over dt} = \Phi_{\mathrm{N}_2}{k_{12a}N_{\mathrm{CH}_4} \over \sum_j k_{1j} N_j } .
\end{equation}
The H$_2$CN radical leads to HCN.
Reaction paths through NO and HNO end in reactions with N that reconstitute N$_2$.  
These are the most important paths when CO$_2$ is abundant.
The NH radical can be important when H$_2$ is very abundant, but under these circumstances
NH is more likely to be recycled to N$_2$ than to react with CH$_n$ to form C-N bonds.
Ground state N
 reacts quickly with CH$_n$ to make HCN, or if CH$_n$ is not abundant it can  
be recycled to N$_2$ through reactions with NH or NO, or following reaction with OH. 
For most cases of interest here, OH is strongly suppressed by reactions with abundant CO or H$_2$.

Ammonia can be abundant after some impacts. As a placeholder, we assume that it is photolyzed. 
\begin{equation}
\label{dNH3dt}
 {dN_{\mathrm{NH}_3} \over dt} = -\Phi_{\mathrm{NH}_3} .
\end{equation}
If methane is also abundant, ammonia photolysis
probably leads to amines, but if H$_2$ is more abundant, ammonia photolysis will mostly end with reconstitution of N$_2$.
The fraction of NH$_3$ photolyses that lead to amines or nitriles is approximated by
\begin{equation}
\label{f_99}
f_{99} = { {dN_{\mathrm{CH}_4} / dt} \over {dN_{\mathrm{CH}_4}/dt} + \Phi_{\mathrm{H}_2\mathrm{O}} + \Phi_{\mathrm{CO}_2} }.
\end{equation}
Reactions of NH$_n$ with H$_2$ will reconstitute NH$_3$ and can be ignored.
Efficient formation of cyanamide (NH$_2$CN) may require NH$_3$.

Net HCN, nitrile, and amine production is approximated by
\begin{equation}
\label{dHCNdt}
 {dN_{\mathrm{HCN}} \over dt} = 2\Phi_{\mathrm{N}_2}f_{1x}  
 \left( d N_{\mathrm{CH}_x}/d t  \over 2\Phi_{\mathrm{N}_2}f_{1x} + d N_{\mathrm{CH}_x}/d t \right)
 + f_{99}\Phi_{\mathrm{NH}_3}.
\end{equation}
where $f_{1x}$ represents the fraction of excited N($^2$D) atoms produced that are available to make HCN,
\begin{equation}
\label{f_1x}
f_{1x} = {k_{12}N_{\mathrm{CH}_4} + k_{15} N_{\mathrm{CO}} +  k_{16}N_{\mathrm{N}_2} \over \sum_j k_{1j} N_j } .
\end{equation}
Equation \ref{dHCNdt} understates the possibility of NH reacting with organic species to make nitriles or amines.
The corresponding net loss of N$_2$ by photolysis is
\begin{equation}
\label{dN2dt}
 {dN_{\mathrm{N}_2} \over dt} = -\Phi_{\mathrm{N}_2} +{1\over 2}{dN_{\mathrm{HCN}} \over dt}
  + {1\over 2}\Phi_{\mathrm{NH}_3} \left( 1 - f_{99}\right) .
\end{equation}
A convenient simplification is that these are reduced atmospheres with no net production of nitrogen oxides.

\medskip
The chief chemical sinks of HCN are addition reactions with OH and H.
The direct reaction with OH has an exothermic branch with products CO and NH$_2$,
but the substantial rearrangements required to get these products require leaping over two energy barriers \citep{Dean2000}.
The climb over the first barrier gives, as one possible set of products, atomic H and HNCO (isocyanic acid).  
Addition reactions with H or CH$_n$ can lead eventually to full hydrogenation
through various intermediates including cyanamide and methlyamine.  Like oxidation,
these paths are expected to be kinetically inhibited.

The important physical sink is rainout.  HCN is not very soluble in water (its Henry Law coefficient is not
very high) but it is miscible.  Total nitrile production is approximated by
\begin{equation}
\label{nitrile}
N_{\mathrm{HCN}} = \int \frac{d N_{\mathrm{HCN}}} {d t} dt .
\end{equation}

\subsubsection{Hydrogen and hydrogen escape}
\label{section:Hydrogen escape}

The most important loss process for hydrogen is escape and its most important sources are 
CH$_4$ photolysis and oxidation, and water photolysis.
\citet{Zahnle2019} found that, for a wide range of solar EUV fluxes and hydrogen mixing ratios,
hydrogen escape from a terrestrial CO$_2$-H$_2$ atmosphere can be approximated by
\begin{equation}
\label{escape}
\left( \frac{d N_{\mathrm{H}_2}}{d t} \right)_{\!\!esc} \approx -\frac{A S_1}{\sqrt{1+B^2S_1^2}}\frac{N_{\mathrm{H}_2}}{\sum_jN_j}
\quad \mathrm{cm}^{-2}\mathrm{s}^{-1}.
\end{equation}
where $A=2\times 10^{12} \mathrm{cm}^{-2}\mathrm{s}^{-1}$ and $B^2=0.006$.
Here we use $S_1$ from Table \ref{tab:EUV} to scale EUV radiation to the levels appropriate to the young Sun.
Equation \ref{escape} blends the energy-limited escape (the limit where $S_1$ is small)
with the diffusion-limited escape (the limit where $S_1$ is large).
Photochemical destruction of H$_2$ is not a concern for the hydrogen budget because 
in the diffusion limit H and H$_2$ escape almost equally easily.


Equation \ref{escape} is readily generalized to other planets and other atmospheric compostions by recognizing
that $A$ is proportional to the density of the planet and $A\div B = b_{ia} \left( H_a^{-1}-H_{\mathrm{H}_2}^{-1}\right)$, 
where $H_{\mathrm{H}_2}$ and $H_a$ represent the unperturbed scale heights of H$_2$ and the background static atmosphere
at the homopause, and $b_{ia}$ represents the binary diffusivity between H$_2$ and the background atmosphere.
The latter is roughly the same for CO$_2$, CO, N$_2$, and CH$_4$ \citep{Marrero1972}.
We can ignore $H_{\mathrm{H}_2}$ at this level of approximation.
Thus, for Earth, $B \propto m_a\div m_{\mathrm{CO}_2}$, where the mean molecular mass of the static gases is
%
\begin{equation}
m_a = { \sum_j N_jm_j - N_{\mathrm{H}_2}m_{\mathrm{H}_2}  \over  \sum_j N_j - N_{\mathrm{H}_2} }
\end{equation}
Other things equal, the diffusion-limited hydrogen escape rate is about three times greater in CO$_2$ than in CH$_4$.
We take the variation of $B$ as a function of $m_a$ into account in our models.

\medskip

Sources of H$_2$ are photochemical or geological.
The direct source is methane: each methane lost creates the equivalent of two H$_2$ molecules.
Another source of H$_2$ is the water that oxidizes carbon from CH$_4$ to CO and CO$_2$. 
In evaluating this source we hold H$_2$O constant.  The presumption is that water is in equilibrium with an ocean
and resupplied to the stratosphere as needed. 
Some hydrogen is removed from the atmosphere when it is incorporated in precipitating organics and nitriles.
For these we assume an H/C ratio of unity. 
The total rate of change of hydrogen is then
\begin{equation}
\label{dH2dt}
\frac{d N_{\mathrm{H}_2}}{dt} = \left( \frac{d N_{\mathrm{H}_2}}{dt} \right)_{\!\!esc} - 2{dN_{\mathrm{CH}_4}\over dt}
+  2{dN_{\mathrm{CO}_2}\over dt} +  {dN_{\mathrm{CO}}\over dt}
- {dN_{\mathrm{haze}}\over dt} - {dN_{\mathrm{HCN}}\over dt} . 
\end{equation}

\subsection{Photochemical Results}
\label{section:Photochemical Results}

Here we present some illustrative examples of photochemical evolution for impacts of several scales.

 \subsubsection{Vestas}
\label{section:Vesta}

\begin{figure}[!tbh]
 \centering
    \includegraphics[width=0.9\textwidth]{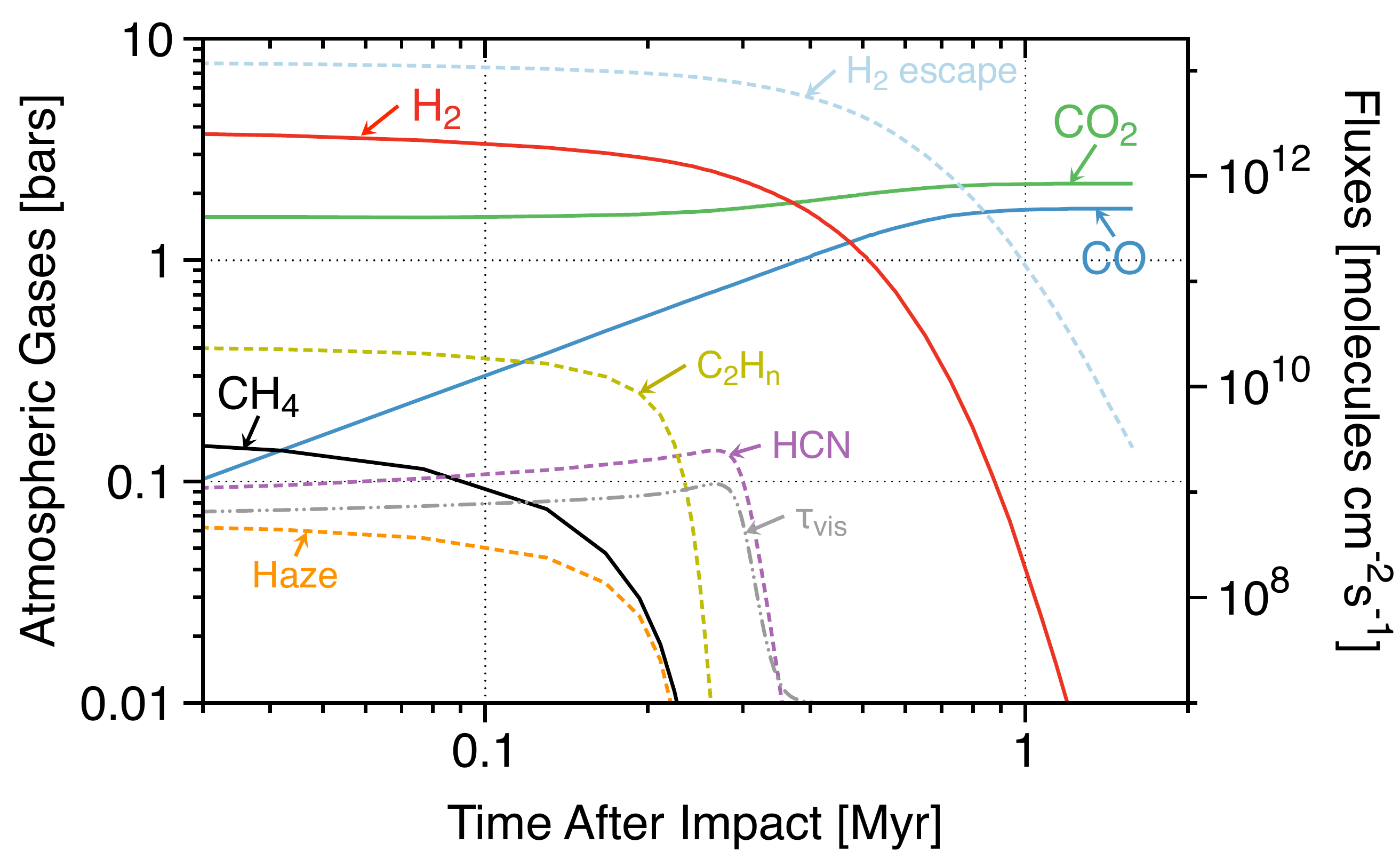} 
\caption{Photochemical dissipation of an atmosphere after the impact of a Vesta-sized body.
The pre-impact atmosphere contained 5 bars of CO$_2$ and 1 bar of N$_2$ over 1.85 oceans (500 bars) of H$_2$O.
This example assumes a dry ($1$ ppm H$_2$O) stratosphere.
The optical depth of organic hazes at 564 nm $\tau_{vis}$ is shown against the left-hand axis.
Production rates of HCN, C$_2$H$_n$ species, and haze are plotted against the right-hand axis.
The H$_2$ escape flux is also plotted against the right-hand axis.
Note that in this case nitrile production (``HCN'') continues well after the other hydrocarbons have dissipated; the 
residual optical depth --- of order unity at 250 nm --- is from nitrogenous hazes.
}
 \label{dry_Vesta}
\end{figure}

A Vesta-scale impact is at the upper limit of what life might survive or prebiotic biomolecules might survive.
Vesta itself is 525 km diameter, and has about 1\% of the mass of the entire late veneer.
There is energy enough to evaporate two oceans of water, which leaves few refugia unless the oceans
were comparably enlarged.

Figure \ref{dry_Vesta} shows evolution after a Vesta-size impact
into a hefty pre-impact atmosphere holding 5 bars of CO$_2$ and 1 bar of N$_2$,
and 1.85 oceans (500 bars) of H$_2$O.
If all the Fe is used, the impact creates 3.9 bars of H$_2$ and converts about 9\% of the 5 bars of CO$_2$ into CH$_4$ (see Table 1).
Subsequent photochemical evolution assumes 1 ppm H$_2$O in the stratosphere, slightly drier than modern Earth's.
Production rates of HCN, C$_2$H$_n$ species, and haze are roughly $30\times$ larger than those on modern Titan, or
comparable to modern volcanic emissions of SO$_2$ or modern lightning production of NO. 
Cumulative precipitation of organics is about 5 cm.    
Hydrogen equivalent to 45 meters of a global ocean escapes to space over the course of the event.

\begin{figure}[!tbh]
 \centering
    \includegraphics[width=0.9\textwidth]{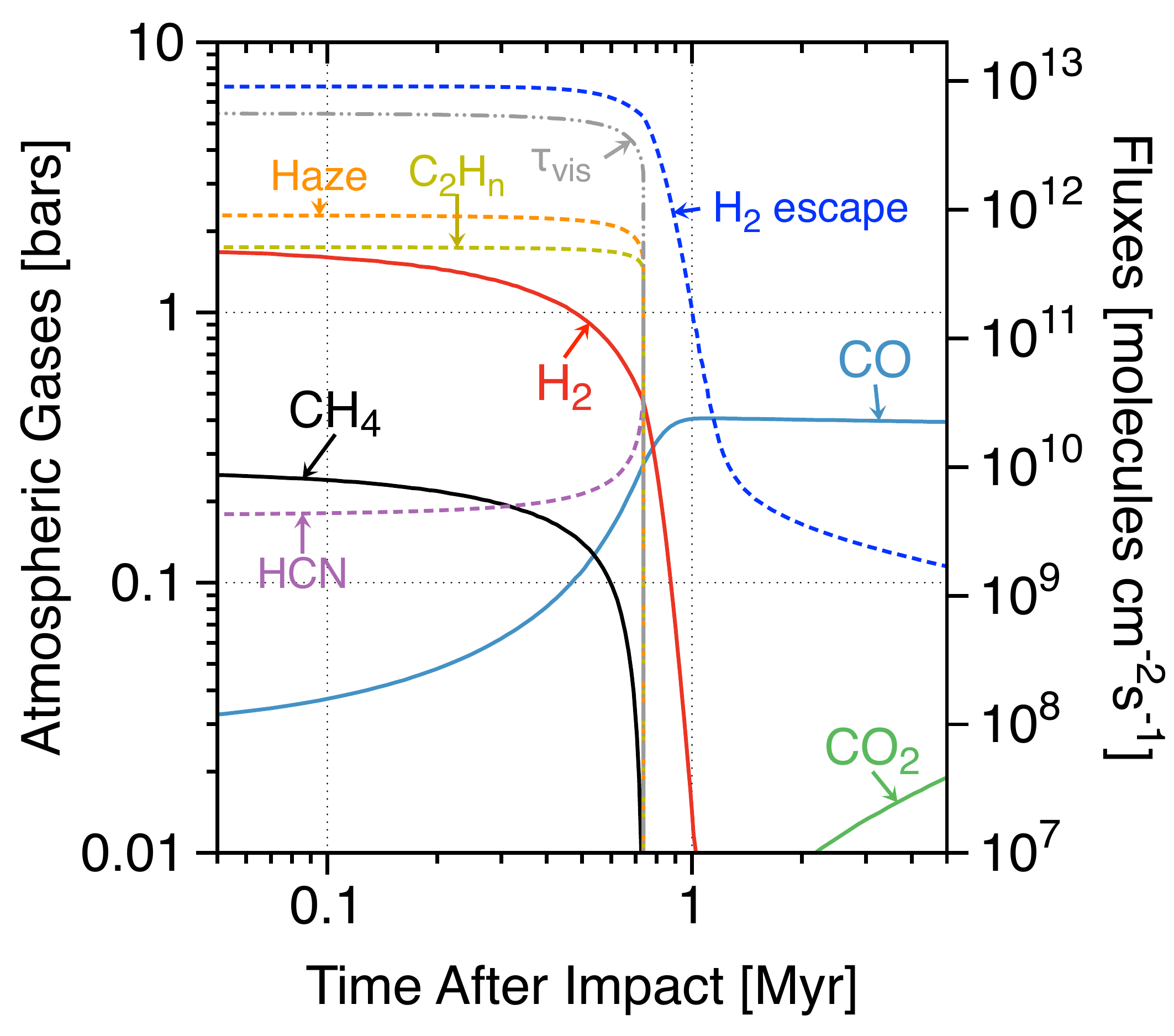} 
\caption{Photochemical evolution after a Vesta-size impact assuming that the atmosphere and ocean 
 equilibrate with the QFM buffer at 650 K (see Figure \ref{QFM_crust_C2_T650}).
The pre-impact atmosphere contained 2 bars of CO$_2$, otherwise conditions are the same
as for Figure \ref{dry_Vesta}.  The case is listed in Table 1.
}
 \label{QFM650_Vesta}
\end{figure}

We estimate the surface temperature by assuming 
that the troposphere follows a moist adiabat, with the tropopause at the skin temperature,
 \begin{equation}
 \label{temperature}
 T_{\mathrm{surf}} \approx \left( \left(1-A\right) SF_{\odot} \over 8\sigma_B \right)^{0.25} \left(p_{\mathrm{surf}} \over p_{\mathrm{tr}}\right)^{(\gamma-1)/\gamma} = 180 \left(p_{\mathrm{surf}} \over 0.1\;\mathrm{bar} \right)^{0.13} \,\mathrm{K} .
 \end{equation}
  We put the tropopause at 0.1 bar as in most solar system planets with atmospheres \citep{Robinson2014}. 
  The solar constant $F_{\odot}=1.36 \times 10^6$ ergs cm$^{-2}$s$^{-1}$
  and the Stefan-Boltzmann constant $\sigma_B = 5.67 \times 10^{-5}$ ergs cm$^{-2}$s$^{-1}$ K$^{-4}$.
 Earth's modern surface temperature is recovered with albedo $A=0.3$ and $\gamma=1.15$.  
 For early Earth, the young Sun is 72\% as bright as the modern Sun ($S=0.72$).
 If we take $0.3<A<0.5$, the $\sim 6$ bar atmosphere (1.6 bars CO$_2$, 0.55 bar N$_2$, 3.9 bars H$_2$, 0.17 bar CH$_4$)
implies a surface temperature $T_{\rm surf} \sim 320 $ K.
 A proper radiative-moist-convective model would be required to provide better estimates of surface temperature and tropopause conditions.
 
Figure \ref{QFM650_Vesta}, a more productive scenario than Figure \ref{dry_Vesta}, is obtained if the reducing
power of FeO in pre-existing mantle and crust is exploited by lowering the quench temperature to the critical point of water and imposing
the QFM buffer, as discussed above in the context of Figure \ref{QFM_crust_C2_T650}.
For a Vesta-size impact striking 2 bars of CO$_2$, 1 bar of N$_2$, and 1.85 oceans (500 bars) of H$_2$O,
%
the result is 1.8 bars of H$_2$ and conversion of more than 90\% of the CO$_2$ into CH$_4$ and more than 10\% of the nitrogen into ammonia
(see Table \ref{table_two}).
The subsequent photochemical evolution in Figure \ref{QFM650_Vesta} assumes a dry 0.1 ppm H$_2$O stratosphere.
 Such dryness might be expected in a deep greenhouse atmosphere illuminated by the faint young Sun.  
Predicted production rates of HCN, C$_2$H$_n$ species, and haze are roughly $1000\times$ larger than those on modern Titan
for about 0.7 Myrs.  
Cumulative precipitation of organics is about 10 m.    
Hydrogen equivalent to 60 meters of a global ocean escapes to space over the course of the event.

\subsubsection{A ``pretty big'' impact}
\label{section:prettybig}

\begin{figure}[!tbh]
 \centering
    \includegraphics[width=0.9\textwidth]{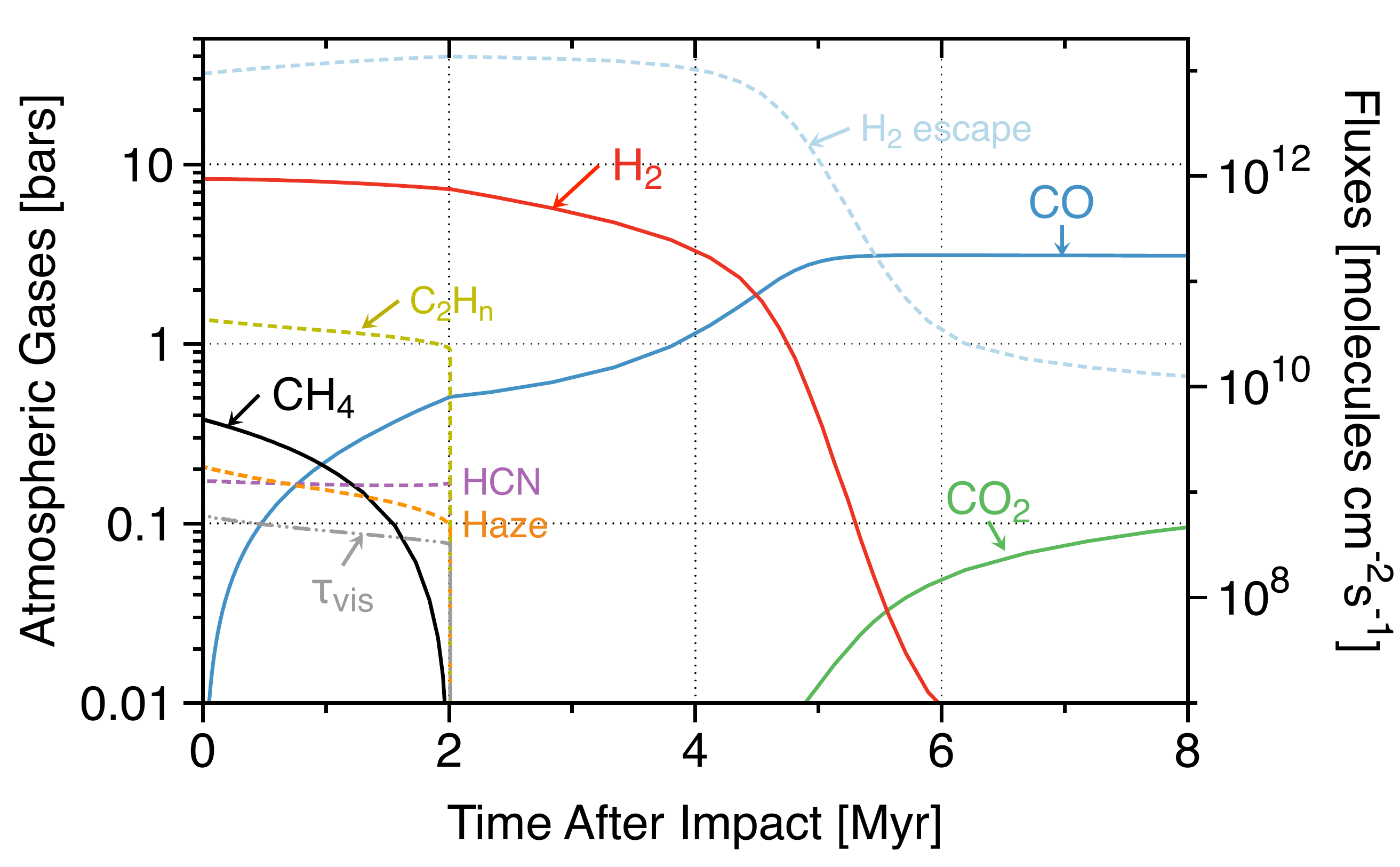} 
\caption{Photochemical dissipation of an atmosphere perturbed by a pretty big $2.5\times 10^{24}$ g impact.
The pre-impact atmosphere contained 5 bars of CO$_2$ and 1 bar of N$_2$ over 1.85 oceans (500 bars) of H$_2$O.
Partial pressures are shown against the left-hand axis.
This example assumes $1$ ppm H$_2$O in the stratosphere, like Earth today. 
Production rates of hazes, C$_2$H$_n$ species, and HCN are indicated against the right-hand axis.
The H$_2$ escape flux is also plotted against the right-hand axis.
Hydrogen from 50 bars (0.2 oceans, 500 meters) of H$_2$O escapes over the course of the event.
}
 \label{PrettyBig-1e-6-C5}
\end{figure}

\begin{figure}[!tbh]
 \centering
    \includegraphics[width=0.9\textwidth]{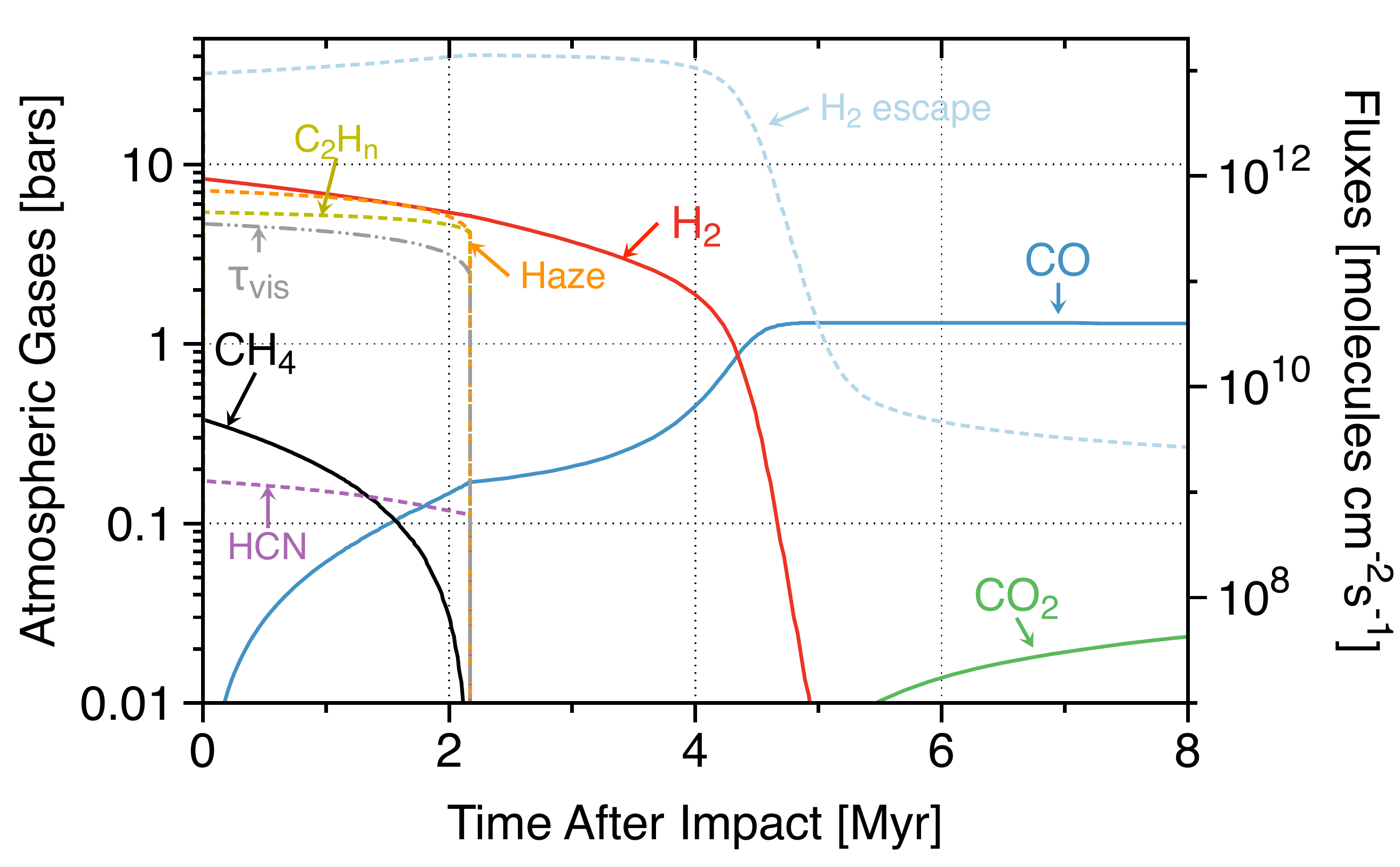} 
\caption{Same as Figure \ref{PrettyBig-1e-6-C5}, but drier (0.1 ppm stratospheric H$_2$O).
 Organic hazes form with optical depths of order 3-5 at 564 nm ($\tau_{vis}$. Haze optical depth is plotted
 against the left-hand axis.
 In this example production rates of HCN and C$_2$H$_n$ species, indicated against the right-hand axis,
are very large, about $500\times$ what they are on modern Titan, or 2\% of modern biotic productivity.
 Cumulative organic precipitation is of the order of 100 moles cm$^{-2}$.
}
 \label{PrettyBig-1e-7-C5}
\end{figure}

Figures \ref{PrettyBig-1e-6-C5} and \ref{PrettyBig-1e-7-C5} document the profound influence of stratospheric moisture
on atmospheric evolution after a bigger impact, here a ``pretty big''
$2.5\times 10^{24}$ g EH-type body striking
a 5 bar CO$_2$, 1 bar N$_2$ atmosphere over 1.85 oceans (500 bars) of liquid H$_2$O.
This approximates the largest event in a minimum late veneer extrapolated from the lunar cratering record, as discussed in Section 2 above.
The surface temperature before the impact may have been in the range $310<T<340$ K (estimated using Eq.\ \ref{temperature}).
There is enough energy released by the impact to vaporize 20 oceans of water or melt the crust to a depth of tens of kilometers. 
It seems unlikely that life on Earth could survive the immediate effects of an impact of this scale,
 but surface conditions ten thousand years later are plausibly temperate enough.
This is an interesting scale for setting the table for life of the future \citep{Benner2019a}.

If all the new iron reacts with water and CO$_2$,
this impact generates 8.4 bars of H$_2$ and converts nearly all of the 5 bars of CO$_2$ into 0.4 bars of CH$_4$.
The mean molecular weight of the air is 3 and the scale height is 100 km. 
These are hydrogen atmospheres resembling Neptune's more than modern Earth's.
Viewed in transit, such an atmosphere would add about 10\% to Earth's apparent diameter, or put another
way, it would lead a distant observer to conclude that Earth had a density of just 4 g cm$^{-3}$.
 
The photochemical evolution in Figure \ref{PrettyBig-1e-6-C5} assumes 1 ppm H$_2$O in the stratosphere. 
The outcome is somewhat similar to that following the Vesta impact, differing mostly in the lack of CO$_2$.
Most of the methane is eventually oxidized. 
A small fraction of the CH$_4$ is built up into organics, nitriles, and hazes.
Production of HCN, hazes, and C$_2$H$_n$ organics is roughly ten times faster 
 than on modern Titan.    
Cumulative precipitation of hazes and nitriles is of the order of half a meter, 
with perhaps another half-meter of partially oxidized organic matter 
(e.g., organic acids and aldehydes stemming from partial oxidation of hydrocarbons).
Hydrogen from 50 bars (0.2 oceans, 500 m) of H$_2$O escapes over 4 Myrs.
Oxidation in the dry stratosphere is too slow to convert CO to CO$_2$ on the timescale of this event.

Figure \ref{PrettyBig-1e-7-C5} is the same impact as Figure \ref{PrettyBig-1e-6-C5},
but evolving with a stratosphere that is ten times drier (0.1 ppm H$_2$O). 
The drier stratosphere might be appropriate given the strong greenhouse effect of the deep troposphere and the faint Sun.
The outcomes are quite different.
The low rates of stratospheric H$_2$O photolysis frustrate oxidation of small organics and thus
allow the buildup of thick photochemical hazes
($\tau_{vis}\sim 5$ at 560 nm according to the fractal model).
 In this example, fully 60\% of the impact-generated CH$_4$ is converted into organic precipitates (hazes),
which correspond in cumulate to a global blanket 10-20 meters thick.  
On the other hand, production of organic nitrogen is no greater because in both case it is limited by the rate of N$_2$ photolysis.
Conditions gradually grow less reducing until the methane is fully titrated and an abrupt bleaching event clears the skies.
In the end, about 40\% of the CH$_4$ is oxidized to CO, and the hydrogen from 0.15 oceans of H$_2$O escape to space.

\subsubsection{The ``maximum HSE'' impact}
\label{section:maximum}

Other things equal, there is a small but significant statistical chance, of the order of 10\%, 
that the last of the world sterilizing events was also the biggest of them. 
Surface environments will not be habitable until well after the impact, but much can be done to prepare the planet for a more hopeful future.
This approximates the impact discussed by \citet{Benner2019a}.


We document two versions of a maximum HSE event, one with a very dry stratosphere and one somewhat moister.
The simulations presume the impact on Earth of a highly-reduced Pluto-sized dwarf planet,
at a time after the Moon-forming impact when there were still 100 bars of CO$_2$ at the surface.
Other initial conditions are 1 bar of N$_2$ and 1.85 oceans of water (500 bars). 
The biggest impact differs from smaller impacts in two key respects: there 
is more iron than CO$_2$ and H$_2$O at the surface, and the ejecta blanket is much deeper than the
oceans.  The former means that the mineral buffer should be important, while the latter hints
that much of the metallic iron might at first be buried.
   
For specificity we presume that the atmosphere and ocean equilibrate with the IW mineral buffer,
and that the remaining metallic iron is oxidized later on geological time scales. 
With these particular assumptions, the impact converts all the CO$_2$ and 15\% of the water to CH$_4$. 
About 60\% of the water (one ocean) remains as H$_2$O.  
Expressed in moles, the atmosphere after the impact contains $6000$ moles H$_2$ per cm$^2$ (40\% 
of the hydrogen in Earth's current oceans) and $2300$ moles CH$_4$ per cm$^2$.
Expressed as pressure, after the impact the dry atmosphere would at first hold 35 bars of H$_2$ and 14 bars of CH$_4$, with a mean
molecular weight of 6.  Because the surface temperature would be high,
a great deal of water would remain in the vapor phase and the actual mean molecular
weight and partial pressures of H$_2$ and CH$_4$ would be correspondingly higher.

 \begin{figure}[!tbh]
 \centering
    \includegraphics[width=0.9\textwidth]{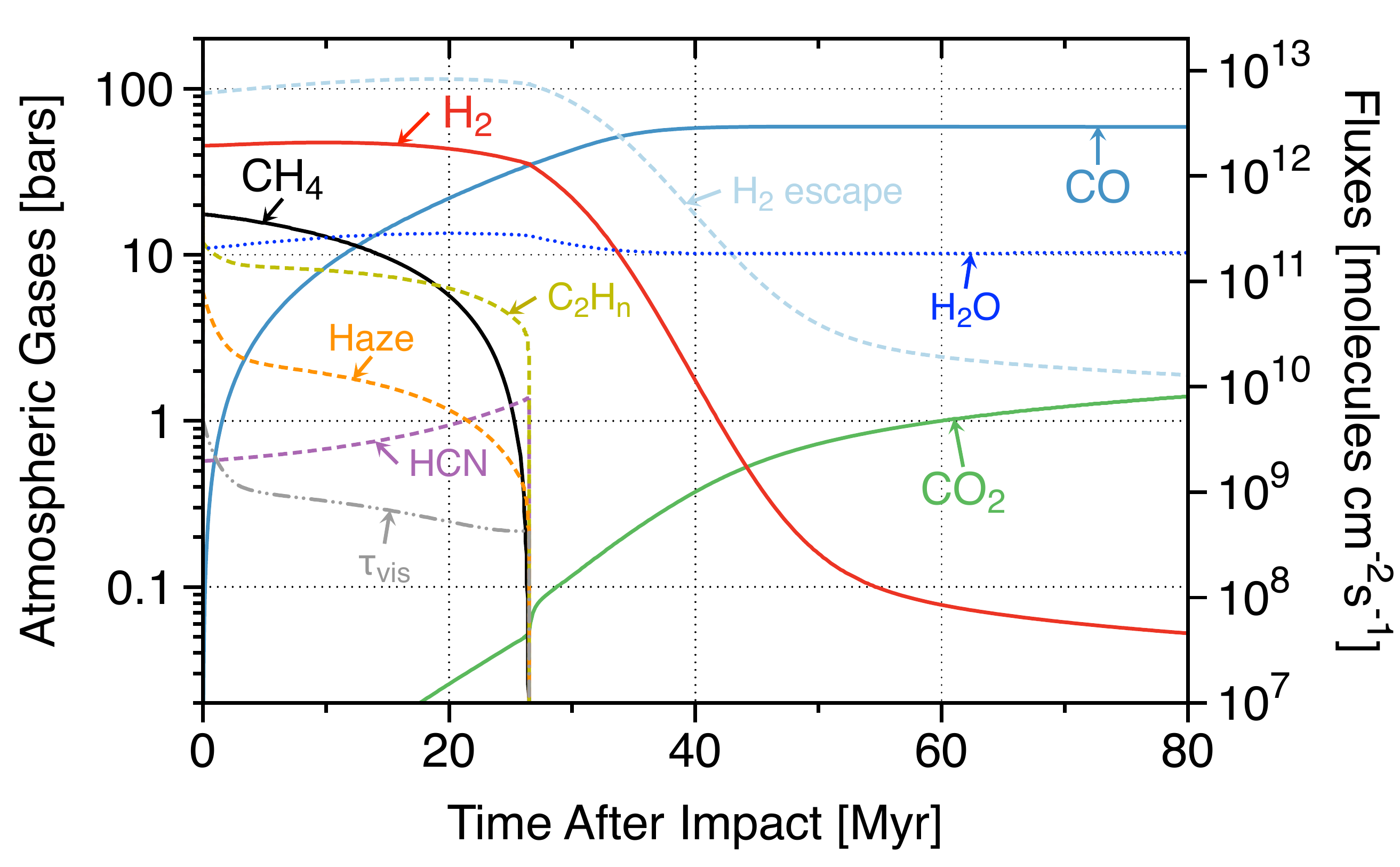} 
\caption{Photochemical evolution of a maximum HSE impact-generated atmosphere 
assuming 1 ppm H$_2$O in the stratosphere.
Initial conditions (100 bars CO$_2$, 1.85 oceans of water) may be appropriate to the largest of Earth's post-Moon impacts.
Partial pressures are shown against the left-hand axis.
Water vapor is mostly confined to the troposphere and presumes a surface temperature governed by 
total atmospheric pressure according to Eq \ref{temperature}. 
Organic C$_2$H$_n$, HCN, and haze production (dashed curves) are mapped to the right hand axis. 
The rate of hydrogen escape is also mapped to the right hand axis. 
Haze optical depth at 560 nm ($\tau_{vis}$) is plotted against the left-hand axis. }
 \label{MAXHSE_1e-6-C100}
\end{figure}

Figure \ref{MAXHSE_1e-6-C100} documents photochemical evolution
with an Earth-like ($1$ ppm H$_2$O) stratosphere.
As with the pretty big impacts, this stratosphere is moist enough that oxidation following
water photolysis is more important than polymerization of CH$_4$.
The photochemical source of organic matter (C$_2$H$_n$, HCN, haze) is nonetheless large,
of the order of $100\times$ that of modern Titan.
This case generates a cumulative global blanket of 25 m of haze organics, plus another 5 m of partially oxidized organics and 5 m of nitrogen-rich organics.  
Much of the nitrogenous material would be amines stemming from impact-generated NH$_3$.
The hydrogen from 2.3 km of water escapes (leaving one ocean of water behind).
Cases with still wetter stratospheres closely resemble this one.

 \begin{figure}[!tbh]
\centering
    \includegraphics[width=0.9\textwidth]{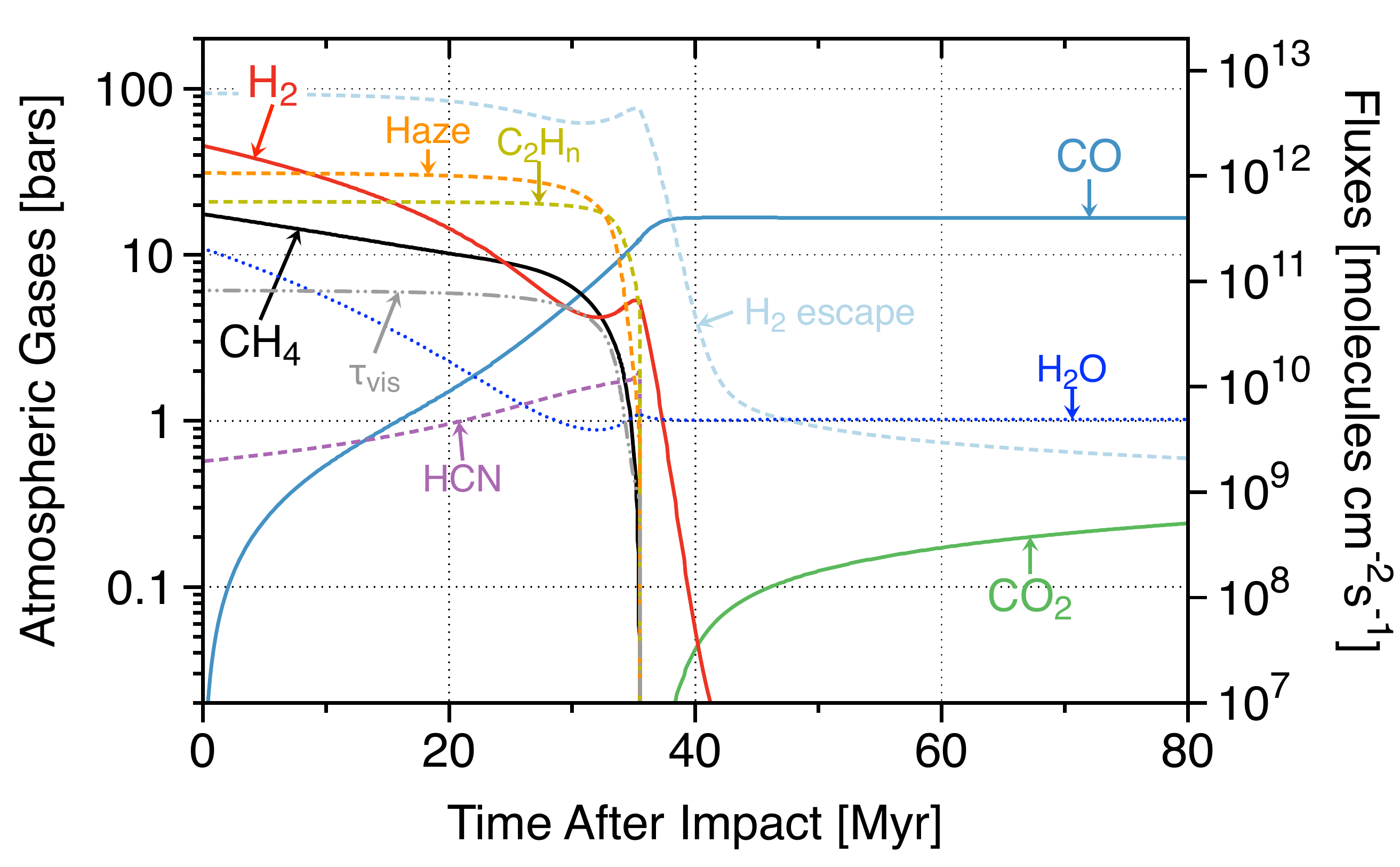} 
\caption{Same as Figure \ref{MAXHSE_1e-6-C100} but with a drier (0.1 ppm H$_2$O) stratosphere.
Water vapor is mostly confined to the troposphere and presumes a surface temperature governed by 
total atmospheric pressure according to Eq \ref{temperature}. 
In this case the evolution is characterized by a very reduced stratosphere and a great deal of Urey-Miller-like 
abiotic organic production that consumes most of the CH$_4$.
Organic (C$_2$H$_n$, HCN, and haze) production rates (dashed curves) are mapped to the right hand axis. 
 Haze optical depth at 560 nm ($\tau_{vis}$) is plotted against the left-hand axis. 
  }
 \label{MAXHSE_1e-7-C100}
 \end{figure}

Figure \ref{MAXHSE_1e-7-C100} documents photochemical evolution with the drier ($0.1$ ppm H$_2$O) stratosphere.
The dry stratosphere leads to highly reduced conditions and rapid photochemical organic haze production at
$10^3\times$ the rate on modern Titan.
Haze optical depths in the UV exceed 100; visible optical depths are in the range of 4-8.
More than 70\% of the methane is polymerized into organic matter equivalent to
a cumulative deposit of the order of 300 to 500 m deep.
In addition, about 10 meters of nitrogenous material reaches the surface,
much of which stems from impact-generated NH$_3$ that is photolyzed in the presence of abundant hydrocarbons,
rather than from photochemical HCN.
The rest of the methane is oxidized to CO.  
The hydrogen from 1.7 km of water escapes.
Cases with even drier stratospheres closely resemble this one.

Despite the thick organic hazes, the climates of the post-impact atmospheres may have been 
 dominated by the greenhouse effect from tens of bars of H$_2$ and CH$_4$ \citep{Wordsworth2013a},
and the surface temperature may have been 350-450 K, or higher.  

\subsubsection{Sub-Vestas}
\label{section:subVesta}

Finally, we consider two impacts that are small enough that life or its precursors ought to survive.  
These have potential to build on what may already have been accomplished.
We look at two cases, the first using the same approach as we used for ocean-vaporizing
impacts, and as this turns out to be rather disappointing, we consider a second case for 
which the assumptions are more liberal and the outcome more bountiful. 

\begin{figure}[!tbh]
 \centering
    \includegraphics[width=0.9\textwidth]{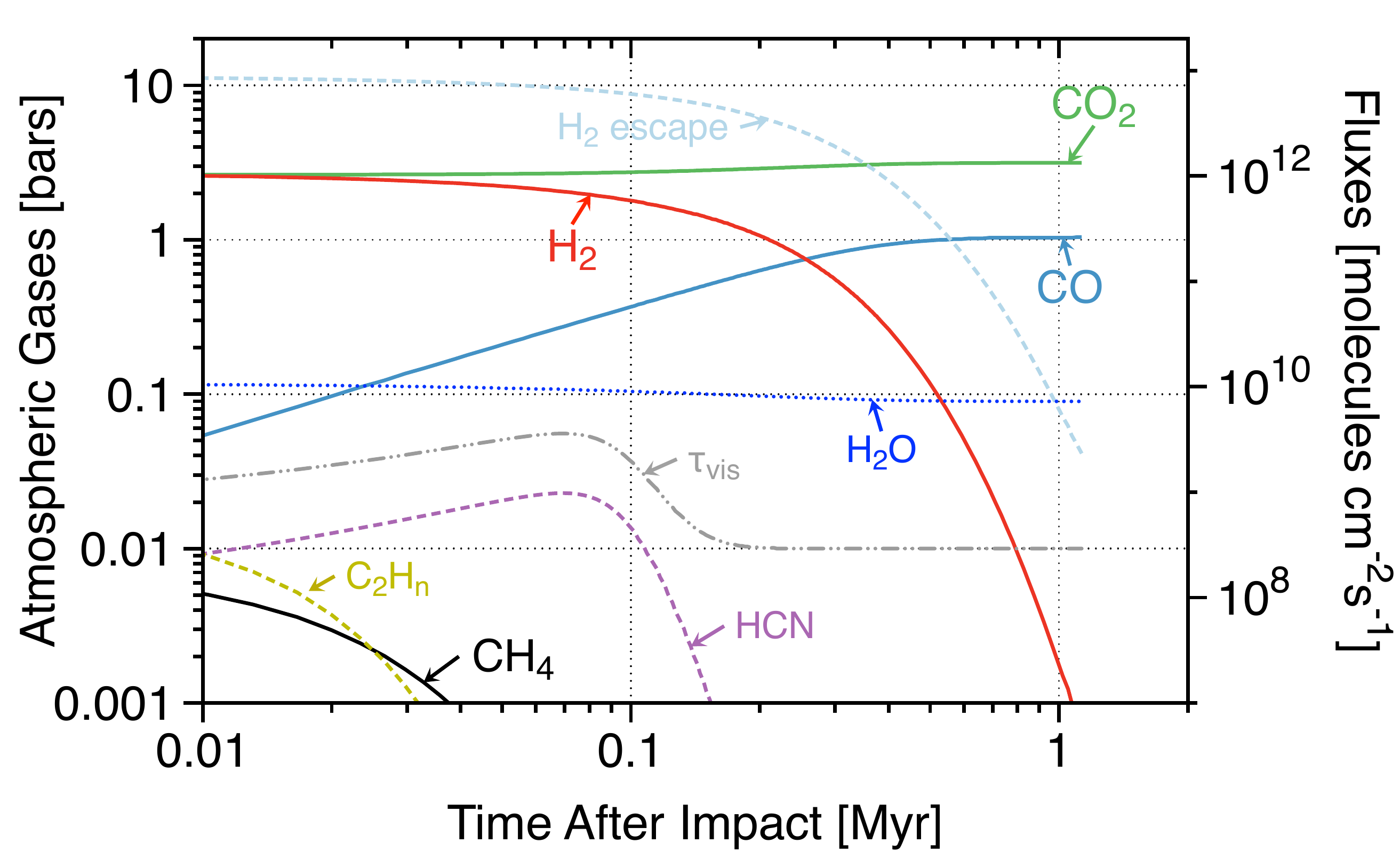} 
\caption{A non-ocean vaporizing impact.  
This case assumes 5 bars CO$_2$ and 1 bar of N$_2$ before the impact.
Quenching is determined by gas phase reactions.
Partial pressures (solid curves) are plotted against the left-hand axis, while fluxes (dashed curves) are plotted 
against the right-hand axis.
Note the prolonged production of HCN in clear skies.
}
 \label{SubVesta_1e-6-C5}
\end{figure}

The first example, Figure \ref{SubVesta_1e-6-C5},
uses only the free Fe of the impact to reduce water and CO$_2$.
This case assumes 5 bars CO$_2$ and 1 bar of N$_2$ in the atmosphere and 1.85 oceans of water
on the surface before the impact.
The impact evaporates about half the ocean, leaving deep waters less disturbed and some subsurface
environments continuously habitable.

The atmosphere immediately after the impact holds 6.2 bars, mostly of H$_2$ and CO$_2$.
This evolves after hydrogen escape into a 5.4 bar CO$_2$-N$_2$-CO atmosphere,
with about 1 bar of CO left at the end of the simulation.
For a short time the atmosphere provides a modest source of nitriles (comparable to modern Titan), but
the cumulative production of organic material over the course of the event is equivalent to just 1 mm of precipitate.
Unlike many of the cases we consider, the results are independent of stratospheric H$_2$O, because CO$_2$ is the oxidant.
The H$_2$ from 15 meters of water escapes.

\begin{figure}[!tbh]
 \centering
    \includegraphics[width=0.9\textwidth]{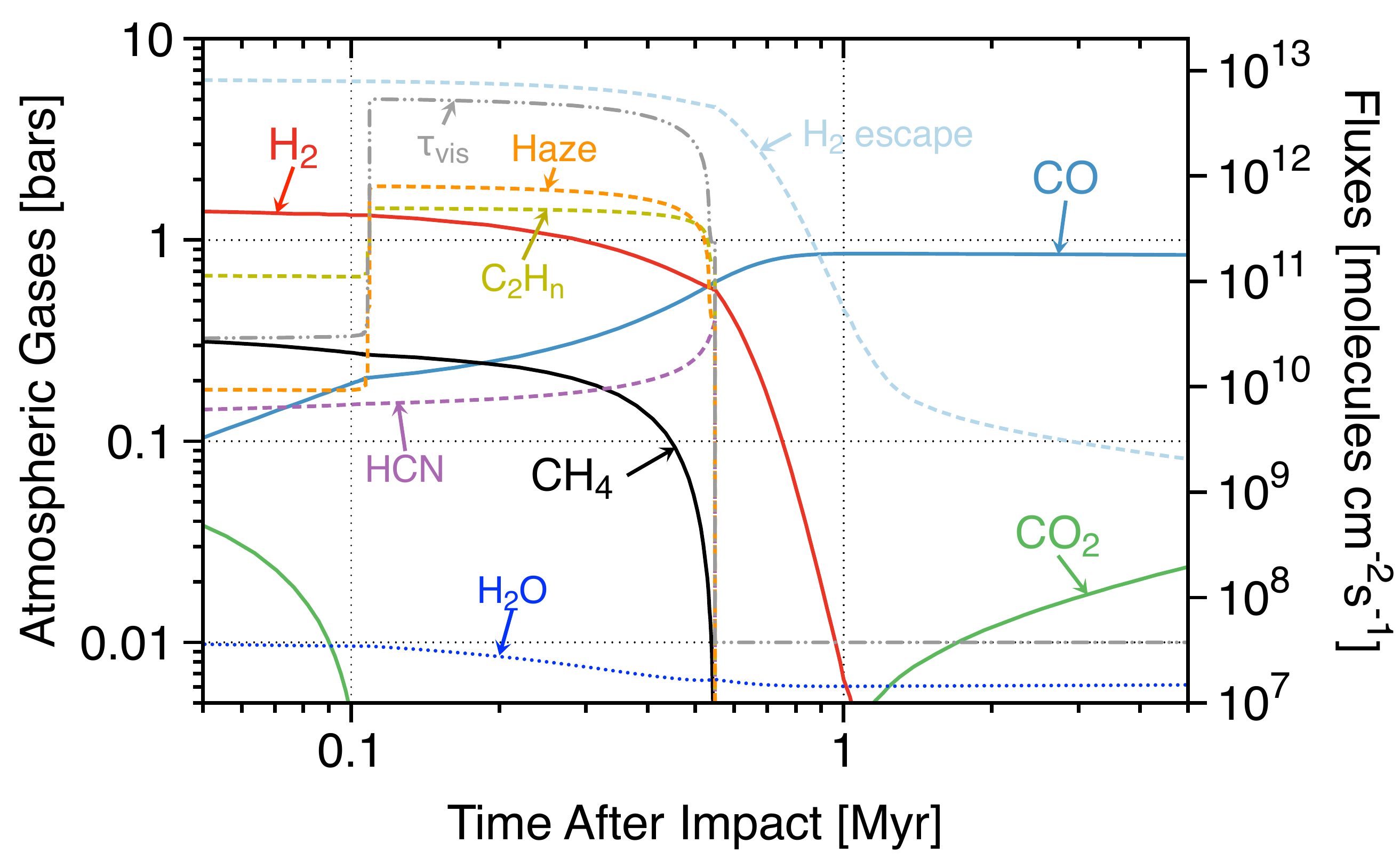} 
\caption{A non-ocean vaporizing impact that assumes that the atmosphere
and ocean equilibrate with the relatively oxidizing QFM buffer at 650 K (i.e., at water's critical temperature). 
Quenching is not determined by gas phase reactions.
This case assumes 2 bars of CO$_2$ and 1 bar of N$_2$ before the impact.
Shown are partial pressures.
}
 \label{SubVesta-1e-7-C2_QFM650}
\end{figure}

The alternative sub-Vesta case (Figure \ref{SubVesta-1e-7-C2_QFM650}) begins with the scenario
presented in Figure \ref{QFM_crust_C2_T650} above, in which the atmosphere and ocean equilibrate with the QFM
buffer at the critical temperature of 650 K.
As in Figure \ref{QFM_crust_C2_T650}, we assume that before the impact the atmosphere held 2 bars of CO$_2$ and one of N$_2$.
The low equilibration temperature favors methane and ammonia,
which are both rather abundant in a 2.3 bar atmosphere volumetrically dominated by 1.5 bars of H$_2$ (see Table \ref{table_two}),
although most of the mass is in CH$_4$, CO$_2$, and N$_2$. 
For the photochemical evolution we assume a dry stratosphere ($0.1$ ppm H$_2$O).  

The story told in Figure \ref{SubVesta-1e-7-C2_QFM650} is eventful.
At first the atmosphere still holds enough CO$_2$ to be weakly oxidizing
and organic production is modest, but after the CO$_2$ is gone stratospheric conditions
become much more reduced, and organic production becomes considerable and stays
so for about 0.5 Myrs.  
This second phase ends abruptly when the methane disappears.
Cumulative production of organics is in the range of 2.5-5 m of hydrocarbons and nitriles.
The H$_2$ from 40 meters of water escapes.
The asymptotic state features about a bar of N$_2$ and CO each, the latter slowly oxidizing to
CO$_2$ on geologic time scales.

\section{Discussion}
\label{section:discussion}

\citet{Benner2019a} have suggested that the greatest of the late veneer impacts
 (corresponding to those we simulate in Section \ref{section:maximum})
created a transiently reducing surface environment on Earth
 ca.\ 4.35 Ga and that this environment lasted for about 15 million years. They also suggest that the origin of the RNA world
 dates to this interval.
They emphasized the difficulty in generating the simple N-rich organic molecules like cyanoacetylene and cyanamide needed to make purines and pyrimidines by paths other than highly reducing atmospheric chemistry.  
   Although the dates and timing of events are less certain than \citet{Benner2019a} seem to suggest, in broad brush we find ourselves
 in general agreement.  The quantity of reducing power is determined from the excess HSEs in the mantle.
 The timescale is set by the EUV radiation emitted by the young Sun.  
 With due allowance for the uncertainty in both, a time scale in the general range of 10-100 Myrs is obtained by simply dividing
 the one quantity by the other, and it cannot be hugely wrong.
 As to when these events took place, opinions can differ, depending on how much weight one places on the ability of zircons to record 
 the evolving conditions at the Earth's surface \citep{Carlson2014}. 
 
 A possible problem posed by the biggest impact is that, although it brings the most reducing power, it is also the most likely 
  to leave the surface too hot to promote prebiotic evolution, and for a long time.
 Even ignoring CO$_2$ and CH$_4$, the greenhouse effect provided by tens of bars of H$_2$ could raise the surface
 temperature to 400 K or more.  Adding CO$_2$ or CH$_4$ or other greenhouse gases would make the surface still hotter. 
 Impacts that are 10- to 100-fold smaller may therefore seem preferable, as these are more likely to
 leave the surface in a temperate state, albeit the reducing conditions do not last as long as for the bigger events.
 In a more recent study, \citet{Benner2019b} concede the potential advantages of less enormous impacts,
 while making a new point: the transience of highly-reduced impact-generated atmospheres
 lets the prebiotic system exploit other chemical
 pathways pertinent to the RNA world that work better in weakly-reduced (QFM mantle-derived) atmospheres.
 These make use of volcanic SO$_2$ (S in a deeply reduced atmosphere would be in H$_2$S),
 borates, and even the highly oxidized Mo$^{+6}$.  
  
 An important opportunity that we have not quantitatively addressed is the delayed oxidation of impactor 
 metallic iron that escaped immediate oxidation by the ocean or atmosphere.
 This iron would instead have been oxidized by recycled surface volatiles over the course
 of geologic time, which may have been millions or many tens of millions of years. 
 If the chief oxidant were water, the chief volcanic gas would have been H$_2$. 
  But if the chief oxidant were CO$_2$
 (probably in the form of subducted carbonate), volcanic gases could have included significant amounts of the much more useful CH$_4$.
 Methane abundances are sensitive to pressure, but to give a specific example,
 CH$_4$ becomes the most abundant C-containing gas at the QFI buffer at 1400 K under 300 bars pressure,
 and it is still 10\% at 1600 K.
 In the latter regime --- one in which CO$_2$ must have been abundant in the atmosphere (else no carbonate)
 and CH$_4$ was abundant in undersea volcanic gases --- 
the chief photochemical products would have been HCN and small organic acids, aldehydes, and carbonyls.
This regime could have lasted for a long time, possibly tens of millions years or more. 
   
\subsection{Ammonia}
\label{section:ammonia}

Large impacts can generate considerable amounts of ammonia from hot H$_2$ and N$_2$.
Ammonia's fate involves many processes that are more complex than what we have been considering in this study.
Nonetheless the inherent interest of ammonia as a constituent of the prebiotic environment is great enough
that we will risk some speculations.    

Ammonia is highly susceptible to UV photolysis at $185<\lambda<215$ nm, where its cross section is of
the order of $3\times 10^{-18}$ cm$^2$, at wavelengths where 
the solar photon flux is much higher than at the shorter
wavelengths where H$_2$O and CO$_2$ absorb.
 The products are NH and NH$_2$ radicals \citep{Huebner1992}
that swiftly react with each other, with oxygen, or with hydrogen in a series of reactions that efficiently recombine N$_2$
\citep{Kuhn1979}. 
Without the protection of high altitude UV absorbers,
a bar of NH$_3$ would revert to N$_2$ in less than $10^4$ years on early Earth.

However, if the atmosphere were also CH$_4$-dominated, as it likely would be were NH$_3$ abundant,
the resulting hydrocarbon hazes might provide UV protection \citep{Sagan1997}.
Perhaps more important is that a significant fraction of the NH and NH$_2$ radicals that photolysis creates
will have good odds of reacting with hydrocarbons to make amines and nitriles. 
Under these conditions the coupled hydrocarbon-ammonia photochemistry would lead to amines and nitriles \citep{Miller1953}.

The other factor to consider with ammonia is that it is very soluble in cool water and hence likely to rain out and partition into the ocean. 
The Henry's Law coefficent for NH$_3$ in water is $K_{H}=4.6\times 10^{-5} e^{-4200/T}$ moles liter$^{-1}$ atm$^{-1}$.
Much of the dissolved NH$_3$ hydrolyzes to ammonium, $\mathrm{NH}_4^{+}$. 
Ammonium abundance is related to NH$_3$(aq) and $pH$ by 
\begin{equation}
\left[\mathrm{NH}_4^{+}\right]\left[\mathrm{OH}^{-}\right]=K_b\left[\mathrm{NH}_3\right]
\end{equation}
in which the base constant $K_b$ is a weak function of temperature \citep{Read1982},
\begin{equation}
 \log{\left(K_b\right)} \approx -4.75-2.5\times 10^{-5}\left(T-298\right). 
 \end{equation}
Expressing $\left[\mathrm{OH}^{-}\right]$ in terms of $pH$ and the auto-ionization constant 
$K_w$ of water 
(which can be crudely approximated as a function of $T$ by $\log{\left(K_w\right)} = -14 +0.03\left(T-298\right) - 7.5\times 10^{-5}\left(T-298\right)^2$),
 the ammonium/ammonia ratio is 
\begin{equation}
\log{\left\{ \left[\mathrm{NH}_4^{+}\right]/\left[\mathrm{NH}_3\right]\right\} }
= \log{\left(K_b\right)}  - \log{\left(K_w\right)} - \log{\left(pH\right)}.
\end{equation}
If all the nitrogen currently in Earth's atmosphere were converted to NH$_3$, the equilibrium NH$_3$ gas pressure above an ocean with $pH$ of 7.8 and temperature 298 K would be just 100 mbars (yet enough to generate a considerable greenhouse effect).
The other 1.56 bars of N would be dissolved in the oceans, 97\% as NH$_4^+$.

Hotter oceans are interesting.  
At 373 K, other things equal, the $\mathrm{NH}_4^+/\mathrm{NH}_3$ ratio would drop to $\sim 2$.
 The equilibrium NH$_3$ gas pressure over the ocean would be 0.02 bars.
The other 1.54 bars of NH$_3$ would be in the oceans, divided between NH$_4^+$ and NH$_3$.
At 500 K the $pH$ would be another unit lower, and NH$_4^+$ less abundant than NH$_3$.
The equilibrium NH$_3$ gas pressure in the atmosphere rises to 0.7-1.0 bar, with the rest of the N divided between
NH$_3$ and NH$_4^+$ dissolved in the ocean.  In this case the ammonia would probably be photochemically destroyed very quickly,
although many of the products would be prebiotically interesting if CH$_4$ were also abundant.

This exercise suggests that NH$_3$ would last longest as ammonium ions in cold seas under an organic haze, conditions
rather similar to those sometimes imagined for early Titan \citep{Lorenz2002}. 
We speculate that ammonia's powerful greenhouse effect can lead to a strong positive
feedback between ocean temperature and exsolution that would in turn speed ammonia's photochemical destruction.
It is not obvious that keeping ammonia around for a long time is a better option for prebiotic purposes than a rapid dump of a wide variety
of nitrogenous products into warm oceans.  Here we simply present these cases as end-members.  

\subsection{The Hadean impact cascade}
\label{section:tall poles}

 \begin{figure}[!tbh]
 \centering
    \includegraphics[width=0.9\textwidth]{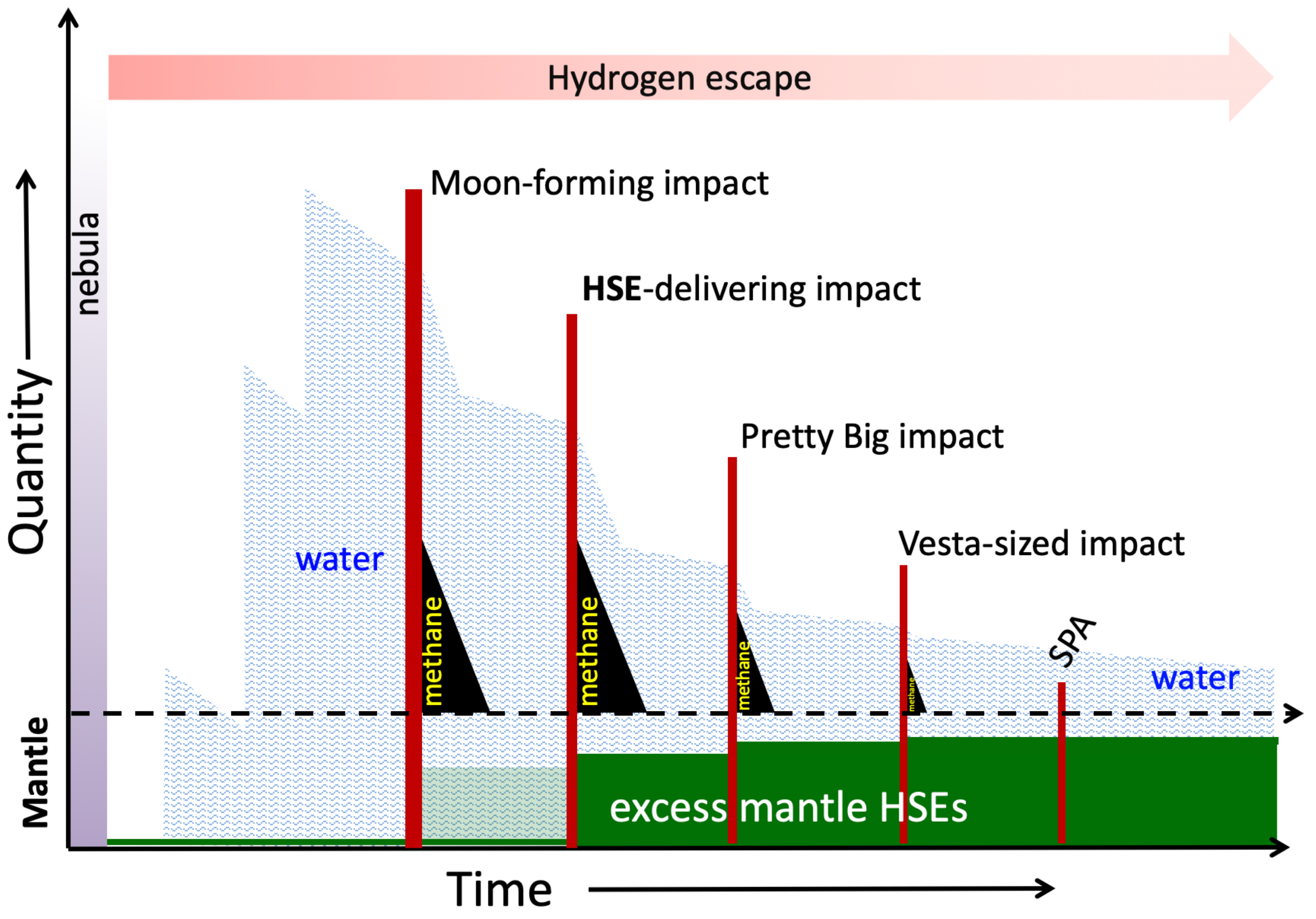}
\caption{A schematic history of water, methane, and highly siderophile elements on Earth in response to late great impacts.
   ``SPA'' refers to an impact on the scale of the one that excavated the lunar S. Pole-Aitken basin --- 
  too small to evaporate the oceans --- which is implied here by submerging SPA's pole under the waves.   
Major impacts are shown decreasing in magnitude as time passed, but the true order leaves much to chance;
it is possible that the largest of the ocean-vaporizers was also the last.}
 \label{Tall poles}
\end{figure}

Figure \ref{Tall poles} presents a notional history of water, methane, and highly siderophile elements (HSEs) on 
early Earth in response to late great impacts. 
 The last stage of this impact history is evident on the face of the Moon and is often called the ``late heavy bombardment,''
often referred to as the ``LHB.'' 
  Water on Earth probably accreted before the Moon-forming impact, and the quantity of water on Earth was likely to have
  decreased during the late accretion of what the isotopic evidence shows were mostly water-poor bodies.  
  Transient methane-rich atmospheres are generated by ocean-vaporizing impacts.
  Here we presume that the mantle's excess HSEs were delivered mostly by a single impact, 
  with light shading indicating an allowance for the HSEs dating to the Moon-forming impact itself.
 Major impacts are shown decreasing in magnitude as time passed, but the true order leaves much to chance;
it is possible that the largest of the ocean-vaporizers was also the last.
Each great impact left Earth in a state resembling somewhat the state of early Earth as sketched by \citet{Urey1952} for periods
 that may have been less than a million years or as long as 100 million years, the duration depending on the size of the impact and on the flux of ultraviolet radiation from the young Sun.  

\section{Conclusions}
\label{section:conclusions}

Great impacts of Earth's late accretion --- especially those that evaporated the oceans --- differ from lesser impacts in several important ways.
First, they delivered significant reducing power in the form of metallic iron to Earth's surface environments.
We infer this because the characteristic isotopic fingerprints 
of the mantle's highly siderophile elements (HSEs) establishes the late veneer as kin to the enstatite chondrites, aubrites, and type IAB iron meteorites, all of which are profoundly reduced and bear ample metallic iron.
The iron was oxidized in the crust or mantle, which we
know because the HSEs are unfractionated, and hence there was no significant loss of metal to the core. 
Water and CO$_2$ were the plausible oxidants; hence hydrogen, carbon monoxide, and methane were the plausible products. 
 Second, the high H$_2$O and H$_2$ vapor pressures in a reduced
 steam atmosphere favor CH$_4$ and NH$_3$ over CO or N$_2$, 
a preference that goes as the square of the pressure.
Third, cooling after impact was slow because the thermal inertia of a hot steam atmosphere containing hundreds of bars of gas is large. 
It takes more than a thousand years to cool 270 bars (an ocean) of steam at the runaway greenhouse cooling rate.
The result is that quench temperatures for gas phase reactions in the CH$_4$-CO-CO$_2$-H$_2$O-H$_2$ system drop to $\sim$800 K or less,
and quench temperatures for the NH$_3$-N$_2$-H$_2$O-H$_2$ system drop to $\sim$1100 K, and hence 
both methane and ammonia form directly by gas phase reactions.  

How much methane actually forms depends on how much carbon was available to the atmosphere before the impact, on how
much reducing power was delivered by the impact,
how much of the delivered iron reacts with the atmosphere and ocean as opposed to being deeply buried under the ejecta,
 and on whether catalysts were active to reduce the quench temperature still further.
In addition to atmospheric CO$_2$ and CO, available carbon inventories would include 
any CO$_2$ dissolved in the oceans, and any CO$_2$ in carbonate rocks that were not buried too deeply to be liberated by shock-heating.
Methane production could have been enormous.
If for example we presume that a maximum-late-veneer scale impact took place on an Earth with a 100 bar of CO$_2$ atmosphere 
\citep[perhaps left over from the Moon-forming impact,][]{Zahnle2007},
and 5 km of water at the surface, there is enough time and reducing power to
convert all of the carbon to methane, 
diluted in several tens of bars of H$_2$. 
The resulting atmosphere is rather Neptune-like, with a scale height in the range of 50-100 km.

Another interesting aspect of the biggest impacts is that much of the delivered iron may have been too deeply buried in the ejecta blanket to be
oxidized in the first thousands of years after the impact.  Under these circumstances the iron must have been oxidized on a longer time
scale set by broadly geological processes that govern the interchange of surface volatiles with crustal and mantle materials. 
Although we have not presented models of such a scenario here, we can expect that gases emitted from a mantle with extant metallic iron would 
be strongly reduced over an extended period of time.  How important this impact coda might be to the origin of life depends on
whether these gases included methane.

Earth's hydrogen-methane atmospheres would have been physically stable --- they did not blow off --- but they were subject to  
photochemical dissipation, as both hydrogen escape and methane photolysis are effectively irreversible.  
The rate the atmosphere evolves is set by the flux of solar far and extreme ultraviolet radiations.
Here we use a zero-D photochemical model to simulate atmospheric evolution:
we count the photons and apportion their effects.
We find that the biggest late veneer impacts can generate (cumulatively) as much as 500 meters of organic over tens of millions of years,
while smaller impacts do commensurately less in commensurately less time.

The details of atmospheric evolution of the transient reduced atmosphere
 are mostly determined by quantities (i.e., bigger impacts have bigger impacts),
but there is one ill-constrained modeling parameter --- the stratospheric water vapor mixing ratio --- to which the outcomes are sensitive.
Moister stratospheres are relatively oxidizing, while drier stratospheres are profoundly reducing and 
resemble conditions on modern Titan, but sped up by a factor of 1000.   
We see a sharp transition in photochemical products determined by the competition
between oxidation and reduction.  In our models, the transition appears
to take place at stratosphere H$_2$O mixing ratios in the range of 0.1-1 ppm, moisture levels
that are not very different from Earth today. 
 A more sophisticated model might predict a smoother transition, or a different critical water abundance for a sharp transition.
 
 Lesser impacts are of course less impactful.
Tens of smaller, non-ocean-vaporizing impacts will generate significant amounts of H$_2$ and CO but very little CH$_4$ or NH$_3$
 unless catalysts were available to reduce the quench temperature.
 Hydrogen and CO are useful ingredients for a bootstrapping origin of life scenario in which
 the biochemical evolution began by catalyzing the kinds of chemical reactions that build hydrocarbons from CO and H$_2$.
 But if the primary requirements for life are methane, ammonia, HCN, and their photochemical derivatives, only the biggest
 impacts or as-yet-unknown chemistry will do.

From the points of view of the origin of life and biblical metaphors, the great impacts may be double-edged swords. 
In their aftermath they leave Earth primed and ready to start life under a classic Urey-Miller H$_2$-rich, CH$_4$-rich,
 possibly even NH$_3$-rich atmosphere that origin of life theorists have long favored.  
 Unfortunately, at least with the biggest of them, the first act is an attempt to wipe out everything that had been accomplished before.
  What this suggests is that impact-generated transient atmospheres may give a planet only one highly favorable roll of the dice. 
 Smaller, less dangerous impacts can inject significant amounts
 of new H$_2$ or CO into the system that might be capable of building on previous progress,
 but these smaller impacts require unidentified catalysts or unidentified
 chemistry to generate large amounts of CH$_4$ or NH$_3$.
 And even if all such conditions were met, 
 there could not have many such impacts, probably no more than a dozen.

\section{Acknowledgements}
This work was in part supported by the NASA Exobiology Program, grant 80NSSC18K1082.  DCC was supported by the Simons Foundation.  We thank Y.\ Abe, S.\ Benner, R.\ Carlson, S.\ Desch, B.\ Fegley Jr., V.S.\ Meadows, A.\ O'Keefe, L.\ Schaefer, N.H.\ Sleep, J.H.\ Waite.

\clearpage
\newpage
\appendix

\subsection*{Appendix A: Mineral buffers, three of them}

We give three representative mineral buffers here.
The quartz-fayalite-iron (QFI) buffer is the most reduced buffer we consider
\begin{equation}
\refstepcounter{reaction}\tag{R\arabic{reaction}}
2{\rm Fe} + {\rm SiO}_2 + {\rm O}_2 \leftrightarrow {\rm Fe}_2{\rm SiO}_4 .
\end{equation}
The oxygen fugacity of the QFI buffer is approximated by
\begin{equation}
\label{QFI}
f_{\mathrm{O}_2} = 1.962\times 10^{-6} T^{3.443} \exp{\left(-54573/T - 2.073\times 10^5/T^{1.5}\right)} .
\end{equation}
Curve fits are based on literature fits for temperatures between
$900\!<\!T\!<\!1420$ K from O'Neill and Eggins 2002.  
The QFI buffer is representative of many chondritic meteorites \citep{Schaefer2017}.

The iron-w{\"u}stite buffer (IW) is based on a simple reaction
\begin{equation}
\refstepcounter{reaction}\tag{R\arabic{reaction}}
{\rm Fe} + {1\over 2} {\rm O}_2 \leftrightarrow {\rm FeO} .
\end{equation}
The oxygen fugacity of the IW buffer is approximated by
\begin{equation}
\label{IW}
f_{\mathrm{O}_2} = 3.1924\times 10^{-6} T^{2.4952} \exp{\left(-39461/T - 3.221\times 10^5/T^{1.5} \right)}
\end{equation}
The direct reaction of water with iron to make FeO and H$_2$
seems kinetically straightforward and hence the IW buffer seems appropriate
for the interaction of hot iron and water.

The fayalite-magnetite-quartz buffer (abbreviated FMQ or QFM) is representative
of modern volcanic degassing and is regarded as typical of the modern Earth's mantle.
It is the most oxidized mineral buffer that we consider.  
The generic QFM reaction is
\begin{equation}
\refstepcounter{reaction}\tag{R\arabic{reaction}}
3{\rm Fe}_2{\rm SiO}_4 + {\rm O}_2  \leftrightarrow  2 {\rm Fe}_3{\rm O}_4 + 3{\rm SiO}_2  .
\end{equation}
The oxygen fugacity of the QFM buffer can be approximated by
\begin{equation}
\label{QFM}
f_{\mathrm{O}_2} = 3.015\times 10^{-4}T^{3.449}\exp{\left(-53649/T\right)} .
\end{equation}
Oxygen fugacity $f_{\mathrm{O}_2}$ has units of atmospheres --- we treat 
$f_{\mathrm{O}_2}$ as effectively the same as the O$_2$ partial pressure $p{\mathrm{O}_2}$.

\newpage
\subsection*{Appendix B.  Chemical reactions discussed in text}
\vspace{-0.2in}
\begin{table}[htbp]
\small
   \begin{tabular}{@{} llcll @{}} 
\multicolumn{4}{l}{Reaction} & rate at 298 K cm$^3$ s$^{-1}$ \\
     \midrule
 R11 &     $\mathrm{N}(^2\mathrm{D}) + \mathrm{H}_2$ & $\rightarrow$ & $ \mathrm{NH} + \mathrm{H} $  & $k_{11} = 2.2 \times 10^{-12} $\\
 R12 &          $\mathrm{N}(^2\mathrm{D}) + \mathrm{CH}_4$ & $\rightarrow$ & products  & $k_{12} = 4 \times 10^{-12} $ \\
\phantom{R12} &           \phantom{$\mathrm{N}(^2\mathrm{D}) + \mathrm{CH}_4$} & $\rightarrow$ & $ \mathrm{H}_2\mathrm{CNH} + \mathrm{H} $  & $k_{12a} = 0.75 \times k_{12} $ \\
\phantom{R12} &    \phantom{$\mathrm{N}(^2\mathrm{D}) + \mathrm{CH}_4$} & $\rightarrow$ & $  \mathrm{NH} + \mathrm{CH}_3 $  & $k_{12b} = 0.25\times k_{12}$ \\
R13 & $\mathrm{N}(^2\mathrm{D}) + \mathrm{H}_2\mathrm{O}$ & $\rightarrow$ & products$^a$  & $k_{13} = 5 \times 10^{-11} $ \\
R14 & $\mathrm{N}(^2\mathrm{D}) + \mathrm{CO}_2$ & $\rightarrow$ & $  \mathrm{NO} + \mathrm{CO} $  & $k_{14} = 3.6 \times 10^{-13} $ \\
 R15 &  $\mathrm{N}(^2\mathrm{D}) + \mathrm{CO}$ & $\rightarrow$ & $  \mathrm{N} + \mathrm{CO} $  & $k_{15}= 1.9 \times 10^{-12}$ \\
R16 &  $\mathrm{N}(^2\mathrm{D}) + \mathrm{N}_2$ & $\rightarrow $ & $ \mathrm{N} + \mathrm{N}_2 $  & $k_{16} = 1.7 \times 10^{-14} $ \\
      \midrule
 R21 &     $\mathrm{O}(^1\mathrm{D}) + \mathrm{H}_2$ & $\rightarrow$ & $ \mathrm{OH} + \mathrm{H} $  & $k_{21} = 1.1 \times 10^{-10} $\\
 R22 &      $\mathrm{O}(^1\mathrm{D}) + \mathrm{CH}_4$ & $\rightarrow$ & products  & $k_{22} = 1.5 \times 10^{-10} $ \\
\phantom{R22} &           \phantom{$\mathrm{O}(^1\mathrm{D}) + \mathrm{CH}_4$} & $\rightarrow$ & $ \mathrm{H}_2\mathrm{COH} + \mathrm{H} $  & $k_{22a}=0.1 \times k_{22}$ \\
\phantom{R22} &           \phantom{$\mathrm{O}(^1\mathrm{D}) + \mathrm{CH}_4$} & $\rightarrow$ & $  \mathrm{OH} + \mathrm{CH}_3 $  & $k_{22b}=0.9\times k_{22}$ \\
R23 &  $\mathrm{O}(^1\mathrm{D}) + \mathrm{H}_2\mathrm{O}$ & $\rightarrow$ &  $\mathrm{OH} + \mathrm{OH} $ & $k_{23} = 2.2 \times 10^{-10} $ \\
R24 & $\mathrm{O}(^1\mathrm{D}) + \mathrm{CO}_2$ & $\rightarrow$ & $  \mathrm{O} + \mathrm{CO}_2 $  & $k_{24} = 7.4\times 10^{-11} $ \\
 R25 & $\mathrm{O}(^1\mathrm{D}) + \mathrm{CO}$ & $\rightarrow$ & $  \mathrm{O} + \mathrm{CO} $  & $k_{25} = 7 \times 10^{-11}$ \\
R26 & $\mathrm{O}(^1\mathrm{D}) + \mathrm{N}_2$ & $\rightarrow $ & $ \mathrm{O}+ \mathrm{N}_2 $  & $k_{26} = 1.8 \times 10^{-11} $ \\
      \midrule
R31 &    $\mathrm{OH} + \mathrm{H}_2$ & $\rightarrow$ & $ \mathrm{H} + \mathrm{H}_2\mathrm{O}  $  & $k_{31} = 6 \times 10^{-15} $\\
 R32 &    $\mathrm{OH} + \mathrm{CH}_4$ & $\rightarrow$ & $\mathrm{CH}_3 + \mathrm{H}_2\mathrm{O} $ & $k_{32} = 6 \times 10^{-15} $ \\
 R35 &    $\mathrm{OH} + \mathrm{CO}$ & $\rightarrow$ & $  \mathrm{CO}_2 + \mathrm{H} $  & $k_{35} = 1.2 \times 10^{-13}$ \\
 R37 &   $\mathrm{OH} + \mathrm{CH}_2$ & $\rightarrow$ & $ \mathrm{H}_2\mathrm{CO} + \mathrm{H} $  & $k_{37} = 1.2\times 10^{-10} $ \\
 R38 &     $\mathrm{OH} + \mathrm{CH}_3 $ & $\rightarrow$ &  products & $k_{38} = 6 \times 10^{-11} $ \\
      \midrule
R41 &    $\mathrm{O} + \mathrm{H}_2$ & $\rightarrow$ & $ \mathrm{H} + \mathrm{OH}  $  & $k_{41} = 1 \times 10^{-17} $\\
 R42 &    $\mathrm{O} + \mathrm{CH}_4$ & $\rightarrow$ & $\mathrm{CH}_3 + \mathrm{OH} $ & $k_{42} = 7 \times 10^{-18} $ \\
 R45 &    $\mathrm{O} + \mathrm{CO}$ & $\rightarrow$ & $  \mathrm{CO}_2  $  & $k_{45}^b = 4.0 \times 10^{-17}$ \\
 R47 &     $\mathrm{O} + \mathrm{CH}_2 $ & $\rightarrow$ &  $ \mathrm{HCO} + \mathrm{H} $ & $k_{47} = 1.2 \times 10^{-10} $ \\
 R48 &   $\mathrm{O} + \mathrm{CH}_3$ & $\rightarrow$ & $ \mathrm{H}_2\mathrm{CO} + \mathrm{H} $  & $k_{48} = 1.2\times 10^{-10} $ \\
 R57 &   $\mathrm{N} + \mathrm{CH}_2$ & $\rightarrow$ & $ \mathrm{HCN} + \mathrm{H} $  & $k_{57} = 1.2\times 10^{-10} $ \\
 R58 &   $\mathrm{N} + \mathrm{CH}_3$ & $\rightarrow$ & $ \mathrm{H}_2\mathrm{CN} + \mathrm{H} $  & $k_{58} = 1.1\times 10^{-10} $ \\
 R78 &   $\mathrm{CH}_2 + \mathrm{CH}_3$ & $\rightarrow$ & $ \mathrm{C}_2\mathrm{H}_4 + \mathrm{H} $  & $k_{78} = 7\times 10^{-11} $ \\
    \bottomrule
  \multicolumn{5}{l}{$a$ - Products are plausibly HNO and H.} \\
  \multicolumn{5}{l}{$b$ - High pressure limit}\\
  \end{tabular}
   \label{tab:chemistry}
\end{table}

\newpage
\subsection*{Appendix C. Photoionization}

Photoionizing photons (EUV$_1$ and $S_1$ in Table 2) are important for hydrogen escape but
 less important than photolysis for chemistry.
To first approximation, CO and N$_2$ usually survive photoionization intact by charge exchange.
Photoionization of CO$_2$ usually creates CO$^+_2$ ions, which can dissociate to CO after reaction with atomic H or O,
and can charge exchange with CH$_4$ to make CH$_4^+$, leaving CO$_2$ intact while ultimately disintegrating CH$_4$. 
As we are not tracking H or O, we arbitrarily assume that 20\% of CO$_2$ photoionizations generate CO.

\begin{equation}
\left({d N_{\mathrm{CO}_2} \over dt} \right)_{\!\!ions} = -0.2\Phi_{\mathrm{CO}_2}^{\ast} 
\end{equation}

Photoionization of water usually yields H$_2$O$^+$, which reacts with H$_2$ to
make H$_3$O$^+$ and H; with CH$_4$ to make H$_3$O$^+$ and CH$_3$; and with CO to make HCO$^+$ and OH.
Both H$_3$O$^+$ and HCO$^+$ dissociatively recombine.
The $\mathrm{H}_2\mathrm{O}^+ + \mathrm{CH}_4$ channel counts as a loss for CH$_4$,
while the HCO$^+$ channel counts as a source of OH.
For cases of most interest to us here, H$_2$O will be condensed at the surface and will not be a major constituent
at the top of the atmosphere, the H$_2$O photoionization terms will be small.

The chief chemical consequence of photo-ionization is that CH$_4$ is broken down into reactive CH$_n$ radicals.
Direct photo-ionization mostly yields CH$_4^+$ or CH$_3^+$, which in subsequent reactions are almost
guaranteed to lead to the loss CH$_4$. 
The other molecular ions also react with CH$_4$, sparking chains of reactions that lead to CH$_n$ radicals.
 E.g., N$_2^+$ can exchange with CO to make CO$^+$, CO$^+$ can exchange with CO$_2$ to make CO$_2^+$, CO$_2^+$ can react with CH$_4$ to free CH$_n$ radicals.
The net
\[
\left({d N_{\mathrm{CH}_4} \over dt} \right)_{\!\!ions} = -\Phi_{\mathrm{CH}_4}^{\!\ast} 
-\Phi_{\mathrm{H}_2}^{\!\ast} {N_{\mathrm{CH}_4}\over X_1} 
-\Phi_{\mathrm{CO}_2}^{\!\ast} {N_{\mathrm{CH}_4}\over X_3} 
-\Phi_{\mathrm{CO}}^{\!\ast} \left( {N_{\mathrm{CH}_4}\over X_2} + {N_{\mathrm{CO}_2}\over X_2}{N_{\mathrm{CH}_4}\over X_3} \right)
\]
\begin{equation}
\label{dCH4dt_ions}
-\Phi_{\mathrm{N}_2}^{\!\ast} \left( {N_{\mathrm{CH}_4}\over X_1} + {N_{\mathrm{CO}}\over X_1} {N_{\mathrm{CH}_4}\over X_2}
+ {N_{\mathrm{CO}}\over X_1}{N_{\mathrm{CO}_2}\over X_2}{N_{\mathrm{CH}_4}\over X_3}
+ {N_{\mathrm{H}_2\mathrm{O}}\over X_1}{N_{\mathrm{CH}_4} \over X_4}  \right)
-\Phi_{\mathrm{H}_2\mathrm{O}}^{\!\ast} {N_{\mathrm{CH}_4}\over X_4} 
\end{equation}
where the $X_k$ factors crudely account for the branching patterns in the ion cascades:
\begin{eqnarray}
X_1 &=& N_{\mathrm{H}_2} + N_{\mathrm{CH}_4} + N_{\mathrm{H}_2\mathrm{O}} + N_{\mathrm{CO}} + N_{\mathrm{CO}_2} \\
X_2 &=& N_{\mathrm{CH}_4} + N_{\mathrm{H}_2\mathrm{O}} + N_{\mathrm{CO}_2} \\
X_3 &=& N_{\mathrm{H}_2} + N_{\mathrm{CH}_4} + N_{\mathrm{H}_2\mathrm{O}} \\
X_4 &=& N_{\mathrm{H}_2} + N_{\mathrm{CH}_4} 
\end{eqnarray}
Equation \ref{dCH4dt_ions} represents about 20\% of total CH$_4$ photolysis.

\end{document}